\documentclass[journal]{IEEEtran}
\usepackage{cite}
\usepackage{graphicx}
\usepackage{subfigure}
\usepackage{algorithm,algorithmic}
\usepackage{threeparttable}
\usepackage{array}
\usepackage{amsmath}
\usepackage{amsfonts,amssymb}
\usepackage{bm}
\usepackage{color}

\usepackage{setspace}

\begin{document}
\title{Statistical Learning Based Joint Antenna Selection and User Scheduling for Single-Cell Massive MIMO Systems}

\author{\IEEEauthorblockN{Mangqing Guo and M. Cenk Gursoy} \\
\thanks{The authors are with the Department of Electrical Engineering and Computer Science, Syracuse University, Syracuse, NY 13244 (email: mguo06@syr.edu, mcgursoy@syr.edu).}
}
\maketitle

\begin{abstract}
Large number of antennas and radio frequency (RF) chains at the base stations (BSs) lead to high energy consumption in massive MIMO systems. Thus, how to improve the energy efficiency (EE) with a computationally efficient approach is a significant challenge in the design of massive MIMO systems. With this motivation, a learning-based stochastic gradient descent algorithm is proposed in this paper to obtain the optimal joint uplink and downlink EE with joint antenna selection and user scheduling in single-cell massive MIMO systems. Using Jensen's inequality and the characteristics of wireless channels, a lower bound on the system throughput is obtained. Subsequently, incorporating the power consumption model, the corresponding lower bound on the EE of the system is identified. Finally, learning-based stochastic gradient descent method is used to solve the joint antenna selection and user scheduling problem, which is a combinatorial optimization problem. Rare event simulation is embedded in the learning-based stochastic gradient descent method to generate samples with very small probabilities. In the analysis, both perfect and imperfect channel side information (CSI) at the BS are considered. Minimum mean-square error (MMSE) channel estimation is employed in the study of the imperfect CSI case. In addition, the effect of a constraint on the number of available RF chains in massive MIMO system is investigated considering both perfect and imperfect CSI at the BS.
\end{abstract}


\section{Introduction}
\IEEEPARstart{M}{assive} MIMO systems have large number of antennas at the base station (BS). Thanks to such large number of antennas, space multiplicity and diversity can be fully used. Indeed, the channels of massive MIMO networks are orthogonal to each other, which can lead to significant growth in the data rates per cell by hundreds of times \cite{Marzett2010,Rana2010}. With such benefits, massive MIMO has become one of the key technologies for 5G cellular systems, and it has been intensively studied in recent years \cite{AndrewsBuzziChoiEtAl2014,LarssonEdforsTufvessonEtAl2014}.

On the other hand, the large number of antennas and RF chains needed at the BS in massive MIMO systems also brings forth challenges regarding the system complexity and hardware energy consumption. Therefore, how to achieve the maximum energy efficiency (EE) is one of the core tasks in the analysis and design of massive MIMO networks. For instance, if uplink and downlink are considered together, improvements in energy efficiency (EE) could be achieved \cite{Bjornson2015}. Also, in order to lower the hardware cost and complexity, the number of RF chains available at the BS can be reduced to be less than the number of antennas. In such cases, how to improve the EE of massive MIMO systems under limitations on the number of RF chains becomes a critical consideration. In addition to just addressing EE, one may also need to consider the tradeoff between the EE and spectral efficiency (SE) in order to achieve the best performance in the system while ensuring a target data transmission level.

While there are various approaches to improve the EE of massive MIMO systems, antenna selection and user scheduling have been commonly considered and studied in this respect. For instance, multi-objective optimization \cite{LiuDuSun2017}, principal component analysis \cite{RanaVesiloCollings2016}, successive removal \cite{BenmimouneDriouchAjibEtAl2015}, ${L_{1/2}}$-regularity based methods \cite{QinLiLvEtAl2016}, norm-and-correlation-based selection algorithms \cite{TaiChungLee2015}, average absolute value of the channel coefficients based methods \cite{ArashYazdianFazelEtAl2017} and convex optimization \cite{GaoEdforsLiuEtAl2013} have been employed in antenna selection to maximize the downlink EE or SE of massive MIMO systems. We note that the number of available RF chains at the BS is limited in \cite{QinLiLvEtAl2016} and \cite{TaiChungLee2015}. Both uplink and downlink are considered and an antenna selection method that searches for the channels with the strongest absolute channel coefficients is applied to maximize the EE of single-cell massive MIMO systems in \cite{ArashYazdianFazel2016}. \cite{Lee2013} shows that simple random antenna selection can lead to significant EE gains. Moreover, when the EE-optimal number of antennas at the BS is larger than a certain threshold, then the performance of random antenna selection is already very close to that of the optimum antenna selection. Convex relaxation and greedy approach have been used for antenna selection in \cite{ElkhalilKammounAl-NaffouriEtAl2016} to maximize the sum rate of an uplink single-cell massive MIMO system.

${K^*}$-random user selection, ${K^*}$-location-dependent user selection \cite{LiuGaoYangEtAl2017}, greedy user selection with linear precoding scheme \cite{AlyamiKostanic2016}, and joint antenna selection and user scheduling method \cite{Maimaiti2019} have been used in downlink massive MIMO systems to maximize the sum rate. Greedy two-step joint antenna selection and user scheduling have been used in \cite{DongTangShenzhen2017} to maximize the sum rate of uplink massive MIMO systems, while \cite{BenmimouneDriouchAjibEtAl2015a} has the objective to maximize the downlink sum rate. Norm-based, greedy-based and TCB (throughput and complexity balanced) based sub-optimal iterative joint antenna selection and user scheduling algorithms have been proposed to improve the downlink channel capacity of distributed massive MIMO systems in \cite{XuLiuJiangEtAl2014}. Besides, semidefinite programming (SDP) approach is used in \cite{Rana2017} for state estimation in smart grids.

The authors in \cite{Bjornson2015} work on the EE maximization problem with joint antenna selection and user scheduling. By utilizing the power allocation technique from \cite{PillaiSuelSeunghunCha2005}, each user could achieve the same rate. Then, the antenna selection and user scheduling problem becomes finding the optimal number of antennas and users which could maximize the EE of the system. However, in order to achieve the same rate, more power should be allocated to the user with worse channel, which will intuitively reduce the EE of the system. With this motivation, we work on further improving the EE of massive MIMO systems with joint antenna selection and user scheduling in this paper, based on the learning framework proposed in \cite{Berny2001}. Besides, we also study the effect of the number of available RF chains to the EE of the system.

Machine learning has attracted much interest recently as a promising approach to support smart radio terminals with applications in 5G networks, including cognitive radio systems, massive MIMO, femtocells, heterogeneous networks, smart grid, energy harvesting systems, device-to-device communications, and so on \cite{JiangZhangRenEtAl2017}. Machine learning also provides efficient tools for tackling certain types of non-convex optimization problems. For instance, machine learning is used for user scheduling in \cite{Shi2018} to maximize the sum rate of massive MIMO systems. We pose a combinatorial optimization problem in this paper, and solve it with the statistical reinforcement learning framework proposed in \cite{Berny2001}.

As one of our key contributions in this paper, we propose a learning-based stochastic gradient descent algorithm to obtain the optimal joint uplink and downlink EE of single-cell massive MIMO systems with joint antenna selection and user scheduling, under a limitation on the number of available RF chains. With Jensen's inequality and the power consumption model, the original joint antenna selection and user scheduling problem is converted into a combinatorial optimization problem. The learning-based stochastic gradient descent algorithm proposed in this paper to solve the corresponding combinatorial optimization problem is based on the learning framework proposed in \cite{Berny2001}. However, the original learning framework for combinatorial optimization problem does not converge to the optimum value of the objective function, and it is difficult to generate random samples based on the given distribution parameters directly when the constraints for the problem are strict. By generating a population of $N_1$ samples instead of only one sample and selecting the fittest one at each iteration, and using an efficient subset selection method for rare event simulation proposed in \cite{AuBeck2001}, we overcome the disadvantages of the original learning algorithm, and devise an efficient learning-based stochastic gradient descent algorithm for the joint antenna selection and user scheduling problem considered in this paper.

The organization of this remainder of the paper is as follows. The system model is described in Section \uppercase\expandafter{\romannumeral2}. Then, linear processing under perfect CSI and imperfect CSI in single-cell massive MIMO systems is discussed in Section \uppercase\expandafter{\romannumeral3}. Section \uppercase\expandafter{\romannumeral4} introduces the power consumption model and the energy-efficiency maximization problem. The original learning method for combinatorial optimization problems and our extensions are discussed in Section \uppercase\expandafter{\romannumeral5}. The energy-efficiency maximization algorithm with a limitation on the number of RF chains is provided in Section \uppercase\expandafter{\romannumeral6}. Finally, numerical results are given in Section \uppercase\expandafter{\romannumeral7} and concluding remarks are provided in Section \uppercase\expandafter{\romannumeral8}.

\section{System Model}
\begin{figure}[htbp]
  \centering
  \includegraphics[width=3.3in]{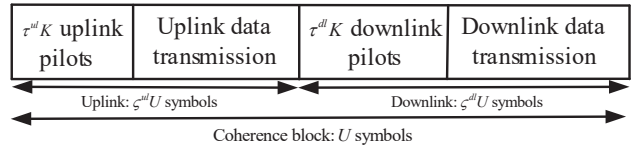}\\
  \caption{Illustration of TDD protocol, where ${\varsigma ^{\textit{ul}}}$ and ${\varsigma ^{\textit{dl}}}$ are the fractions of uplink and downlink transmission, respectively.}\label{fig1}
\end{figure}

Consider a single-cell massive MIMO system consisting of one BS with the antenna set $\cal M$, and $K$ single-antenna users. The user set is denoted by $\cal K $. And we have $|{\cal M}| = M$ and $|{\cal K}| = K$, where $|\cdot|$ denotes the cardinality of a given set. Without loss of generality, we assume that the system operates over a flat-fading channel \footnote{For frequency selective channels, orthogonal frequency division multiplexing (OFDM) can, for instance, be used to create flat-fading subchannels.}, the transmission bandwidth is $B$ Hz, and the channel coherence bandwidth is $B_c$ Hz. $U$ symbols are transmitted during a time-frequency coherence block. The uplink and downlink transmissions are considered together with fixed ratios of ${\varsigma ^{\textit{ul}}}$ and ${\varsigma ^{\textit{dl}}}$, respectively, with ${\varsigma ^{\textit{ul}}}+{\varsigma ^{\textit{dl}}}=1$. During each channel coherence interval, $U{\varsigma ^{\textit{ul}}}$ uplink symbols are transmitted first, then the $U{\varsigma ^{\textit{dl}}}$ downlink symbols. We assume that BS and all users are perfectly synchronized and operating according to the time-division duplex (TDD) protocol. The uplink and downlink channels are considered to be reciprocal and the uplink channel estimation at the BS could be used for both uplink reception and downlink transmission. As shown in Fig. \ref{fig1}, ${\tau ^{\textit{ul}}}K$ pilot symbols are used during uplink transmission for channel estimation, while another ${\tau ^{\textit{dl}}}K$ pilot symbols are used during downlink transmission to estimate each user's effective channel and interference variance under the current precoding \cite{Bjornson2015}. We assume ${\tau ^{\textit{ul}}}$, ${\tau ^{\textit{dl}}} \geqslant 1$ to enable orthogonal pilot sequences among users.

The pilot sequences used by the $K$ users during channel estimation are $\sqrt {{p_{{\textit{pilot}}}}} {{\bm{\Phi }}^T}$, where ${{\bm{\Phi }}^H}{\bm{\Phi }} ={{\tau ^{\textit{ul}}}K}{{\textbf{I}}_K}$ and ${{p_{{\textit{pilot}}}}}$ is the transmit power of pilot signals. Then, the received pilot signal at the BS during each channel coherence interval is
\begin{equation}\label{equ1}
{{\textbf{Y}}_{{\textit{pilot}}}} = \sqrt {{p_{{\textit{pilot}}}}} {\textbf{G}}{{\bm{\Phi}}^T} + {\textbf{N}}
\end{equation}
where ${\textbf{N}}$ is the AWGN noise matrix with i.i.d. complex Gaussian ${\cal C}{\cal N}(0,{\sigma ^2})$ components with zero mean and variance $\sigma^2$, and ${\textbf{G}} = [{{\textbf{g}}_1},{{\textbf{g}}_2},...,{{\textbf{g}}_K}] \in {{\cal C}^{M \times K}}$ is the matrix describing the channel from all the users to the BS. More specifically, ${{\textbf{g}}_k}$ is the uplink channel vector from the $k$th user to all the antennas at the BS. For flat fading channels, ${{\textbf{g}}_k}$ can be written as \cite{Marzett2010},
\begin{equation*}
  {{\textbf{g}}_k} = {\beta _k}{{\textbf{h}}_k}
\end{equation*}
where the vector ${{\textbf{h}}_k}$ of fading coefficients is composed of i.i.d. ${\cal C}{\cal N}(0,1)$ elements. ${\beta _k}$ describes the effects of path loss and shadow fading, and can be expressed as
\begin{equation*}
  {\beta _k} = \frac{{{z_k}}}{{(r_k/d_0)^\gamma }}
\end{equation*}
where $r_k$ is the distance between the $k$th user and the BS, $d_0$ is a reference distance, $\gamma $ is the path loss decay exponent, and the shadow fading coefficient ${{z_k}}$ is a log-normal random variable, i.e., $10{\log _{10}}({z_k})$ is zero-mean Gaussian distributed with standard deviation ${\sigma _{{\rm{\textit{shad}}}}}$.

The received uplink data signal at the BS is
\begin{equation}\label{equ2}
  {{\textbf{y}}^{\textit{ul}}} = \sqrt {{p^{\textit{ul}}}} {\textbf{Gx}} + {\textbf{n}}
\end{equation}
where $\sqrt {{p^{\textit{ul}}}} {\textbf{x}}$ denotes the signal vector transmitted from the users to the BS (with ${\mathbb E}\{ {\textbf{x}}{{\textbf{x}}^H}\}  = {{\textbf{I}}_K}$) and ${\textbf{n}} \sim CN(0,{\sigma ^2}{{\textbf{I}}_M})$ is the AWGN noise vector at the BS.

During downlink data transmission, the received signal vector of all users can be expressed as
\begin{equation}\label{equ3}
  {{\textbf{y}}^{\textit{dl}}} = \sqrt {{p^{\textit{dl}}}} {{\textbf{G}}^H}{\textbf{Wx}} + {\textbf{n}}
\end{equation}
where $\sqrt {{p^{\textit{dl}}}} {\textbf{x}}$ is the signal transmitted from the BS to users, and ${\textbf{W}} = [{{\textbf{w}}_1},{{\textbf{w}}_2},...,{{\textbf{w}}_K}] \in {{\cal C}^{M \times K}}$ is the precoding matrix at the BS.

\section{Linear Processing in Single-Cell Massive MIMO Systems} \label{sec:linearprocessing}
As the number of antennas at the BS is very large in massive MIMO systems, linear precoding schemes can obtain near-optimal performance \cite{RusekPerssonLauEtAl2013}. Therefore, we consider linear processing in this paper. More specifically, we assume that linear processing with maximum ratio combination (MRC) or zero-forcing (ZF) receiver is employed at the BS during uplink data transmission, and maximum ratio transmission (MRT) or ZF precoding is used during downlink transmission, under both perfect and imperfect CSI assumptions. Finally, we obtain lower bounds for the achievable data rates of different users with different linear processing methods.
\subsection{Linear Processing with Perfect CSI}
In this subsection, we assume the availability of perfect CSI at the BS. Both MRC and ZF receiving strategies are considered at the BS during uplink data transmission, while MRT and ZF precoding schemes are considered for downlink data transmission.
\subsubsection{Uplink}
The received uplink signal at the BS after linear processing can be expressed as
\begin{equation}\label{equ4}
  \hat{\textbf{y}}^{\textit{ul}}=\sqrt{p^{\textit{ul}}}\textbf{AGx}+\textbf{An}
\end{equation}
where $\textbf{A}$ is the linear combination matrix at the BS and is given by
\begin{equation}\label{equ5}
  \textbf{A}=
\begin{cases}
   \textbf{G}^H &\text{for MRC}, \\
   (\textbf{G}^H\textbf{G})^{-1}\textbf{G}^H &\text{for ZF}. \\
\end{cases}
\end{equation}

The $k$th component of $\hat{\textbf{y}}^{\textit{ul}}$ can be expressed as
\begin{equation}\label{equ6}
  \hat{\textbf{y}}_{k}^{\textit{ul}}=\sqrt{p^{\textit{ul}}}\textbf{a}_k\textbf{g}_kx_k+\sqrt{p^{\textit{ul}}}\sum\limits_{i = 1,i \ne k}^K {\textbf{a}_k\textbf{g}_ix_i}+\textbf{a}_k\textbf{n}
\end{equation}
where $\textbf{a}_k$ is the $k$th row of $\textbf{A}$. Therefore, the SINR corresponding to $k$th user's signal received at the BS is
\begin{equation}\label{equ7}
  \gamma_{k}^{\textit{ul}}=\frac{p^{\textit{ul}}\left| {\textbf{a}_k\textbf{g}_k} \right|^2}{p^{\textit{ul}}\sum\limits_{i=1, i\ne k}^K\left|{\textbf{a}_k\textbf{g}_i}\right|^2+\left\| {\textbf{a}_k} \right\|^2\sigma^2}.
\end{equation}
Then the achievable uplink data rate (bps/Hz) of $k$th user is
\begin{equation}\label{equ8}
  R_{k}^{\textit{ul}}={\mathbb E}\{\log_2{(1+\gamma_{\textit{p,k}}^{\textit{ul}})}\}.
\end{equation}
Using Jensen's inequality and the characteristics of wireless channels, we obtain the following lower bounds on $R_{k}^{\textit{ul}}$ when MRC and ZF schemes are employed, respectively, at the BS \cite{Ngo2013}:
\begin{equation}\label{equ10}
  \widetilde {R}_{k}^{\textit{MRC,ul}}=\log_2{\left(1+\frac{p^{\textit{ul}}(M-1)\beta_k}{p^{\textit{ul}}\sum\limits_{i=1,i\ne k}^K \beta_i+\sigma^2}\right)},
\end{equation}
\begin{equation}\label{equ11}
  \widetilde {R}_{k}^{\textit{ZF,ul}}=\log_2{\left(1+\frac{p^{\textit{ul}}(M-K)\beta_k}{\sigma^2}\right)}.
\end{equation}
Since the number of antennas at the BS should be no less than the number of users to achieve space multiplicity and diversity gains in MIMO systems, the condition $M\geq K$ should be satisfied in (\ref{equ10}). Because $\mathbb{E}\{ {[{({{\textbf{G}}^{\text{H}}}{\textbf{G}})^{ - 1}}]_{\textit{kk}}}\}  = \frac{1}{{(M - K){\beta _k}}}$, $M\geq K+1$ should be satisfied in (\ref{equ11}).
\subsubsection{Downlink}
We assume MRT or ZF precoding is employed at the BS for downlink transmission and the precoding vector can be expressed as
\begin{equation}\label{equ12}
  \textbf{w}_k=\frac{\textbf{a}_k^H}{\left\| {\textbf{a}_k} \right\|}.
\end{equation}
The received signal at the $k$th user is
\begin{equation}\label{equ13}
  \textbf{y}_{k}^{\textit{dl}}=\sqrt{p^{\textit{dl}}}\textbf{g}_k^H\textbf{w}_kx_k+\sqrt{p^{\textit{dl}}}\sum\limits_{i = 1,i \ne k}^K {\textbf{g}_k^H\textbf{w}_ix_i}+n_k.
\end{equation}
Thus, the SINR at the $k$th user during downlink transmission is
\begin{equation}\label{equ14}
  \gamma_{k}^{\textit{dl}}=\frac{p^{\textit{dl}}\left| {\textbf{g}_k^H\textbf{w}_k} \right|^2}{p^{\textit{dl}}\sum\limits_{i=1, i\ne k}^K\left|{\textbf{g}_k^H\textbf{w}_i}\right|^2+\sigma^2}.
\end{equation}
Then the achievable downlink data rate (bits/s/Hz) of the $k$th user is
\begin{equation}\label{equ15}
  R_{k}^{\textit{dl}}={\mathbb E}\{\log_2{(1+\gamma_{k}^{\textit{dl}})}\}.
\end{equation}
Following the same procedure as in \cite{Ngo2013}, we obtain the following lower bounds on $R_{\textit{p,k}}^{\textit{dl}}$ with MRT and ZF precoding, respectively:
\begin{equation}\label{equ17}
  \widetilde {R}_{k}^{\textit{MRT,dl}}=\log_2{\left(1+\frac{p^{\textit{dl}}(M-1)\beta_k}{p^{\textit{dl}}(K-1)\beta_k+\sigma^2}\right)},
\end{equation}
\begin{equation}\label{equ18}
  \widetilde {R}_{k}^{\textit{ZF,dl}}=\log_2{\left(1+\frac{p^{\textit{dl}}(M-K)\beta_k}{\sigma^2}\right)}.
\end{equation}
Similar to (\ref{equ10}) and (\ref{equ11}), (\ref{equ17}) and (\ref{equ18}) require that the conditions of $M\geq K$ and $M\geq K+1$, respectively, are satisfied. Then, the lower bound on the total achievable data rate in the system is
\begin{IEEEeqnarray}{rCl}\label{eq_t}
  \widetilde {R}&=&\widetilde {R}^{\textit{ul}}+\widetilde {R}^{\textit{dl}} \nonumber\\
  &=&a\sum_{k=1}^{K}\widetilde {R}_{k}^{\textit{ul}}+b\sum_{k=1}^{K}\widetilde {R}_{k}^{\textit{dl}}
\end{IEEEeqnarray}
where
\begin{equation}\label{eq_t1}
  a=\varsigma^{\textit{ul}}\left(1-\frac{{\tau ^{\textit{ul}}} K}{U\varsigma^{\textit{ul}}}\right)B
\end{equation}
\begin{equation}\label{eq_t2}
  b=\varsigma^{\textit{dl}}\left(1-\frac{{\tau ^{\textit{dl}}} K}{U\varsigma^{\textit{dl}}}\right)B
\end{equation}
in which the factor $1-\frac{{\tau ^{\textit{ul}}} K}{U\varsigma^{\textit{ul}}}$ and $1-\frac{{\tau ^{\textit{dl}}} K}{U\varsigma^{\textit{dl}}}$ account for the uplink and downlink pilot overhead, respectively \cite{Bjornson2015}.
\subsection{Linear Processing with Imperfect CSI}
In practice, CSI obtained via channel estimation is imperfect. Let us denote the channel estimate obtained from (\ref{equ1}) as $\hat{\textbf{G}}$. Then, the channel estimation error is
\begin{equation}\label{equ36}
  \bm{\varepsilon}=\hat{\textbf{G}}-\textbf{G}.
\end{equation}
The minimum mean-square error (MMSE) estimate of $\textbf{G}$ given $\textbf{Y}_{{\textit{pilot}}}$ is \cite{BjornsonOttersten2010}
\begin{IEEEeqnarray}{rCl}\label{equ37}
  \widehat {\textbf{G}} = \frac{1}{{\sqrt {{p_{{\textit{pilot}}}}} }}{{\textbf{Y}}_{{\textit{pilot}}}}{{\bm{\Phi }}^ * }{\left( {\frac{{{\sigma ^2}}}{{{p_{{\textit{pilot}}}}}}{{\textbf{D}}^{ - 1}} + {\tau ^{\textit{ul}}}K{{\textbf{I}}_K}} \right)^{ - 1}}
\end{IEEEeqnarray}
where ${{\bm{\Phi }}^*}$ is the complex conjugate of the pilot sequence matrix, and ${\textbf{D}} = {\rm{diag}}({\beta _1},{\beta _2},...,{\beta _K})$. The error covariance
\begin{IEEEeqnarray}{rCl}\label{equ3701}
  {{\textbf{C}}_{\textit{MMSE}}} &=& {\mathbb E}\left\{ {vec({\bm{\varepsilon }}) * vec{{({\bm{\varepsilon }})}^H}} \right\}  \nonumber\\
  &=& {\left( {{{\textbf{D}}^{ - 1}} + \frac{{{\tau ^{\textit{ul}}}K{p_{{\textit{pilot}}}}}}{{{\sigma ^2}}}{{\textbf{I}}_K}} \right)^{ - 1}} \otimes {{\textbf{I}}_M}
\end{IEEEeqnarray}
where $vec( \cdot )$ and $ \otimes $ are matrix vectorization and kronecker product operation, respectively. Therefore, each element of ${\bm{\varepsilon }}$ is independent of each other, and the $k$th column of $\bm{\varepsilon}$, denoted as $\bm{\varepsilon}_k$, is a vector of random variables with zero mean and variance
\begin{equation}\label{eqv2}
  {\gamma _k} = \frac{{{\beta _k}{\sigma ^2}}}{{{\tau ^{\textit{ul}}}K{p_{{\textit{pilot}}}}{\beta _k} + {\sigma ^2}}}, k=1,2,...,K.
\end{equation}
Substituting (\ref{equ1}) into (\ref{equ37}), we can obtain
\begin{equation}\label{equ3702}
  \widehat {\textbf{G}} = \left( {{\tau ^{\textit{ul}}}K{\textbf{G}} + \frac{{{\textbf{N}}{{\bm{\Phi }}^ * }}}{{\sqrt {{p_{{\textit{pilot}}}}} }}} \right){\left( {\frac{{{\sigma ^2}}}{{{p_{{\textit{pilot}}}}}}{{\textbf{D}}^{ - 1}} + {\tau ^{\textit{ul}}}K{{\textbf{I}}_K}} \right)^{ - 1}}.
\end{equation}
Therefore, elements of $\widehat {\textbf{G}}$ are independent of each other, and the $k$th column of $\hat{\textbf{G}}$, denoted as $\hat{\textbf{g}}_k$, is a vector of random variables with zero mean and variance
\begin{equation}\label{eqv1}
  {{\hat \beta }_k} = \frac{{{\tau ^{\textit{ul}}}K{p_{{\textit{pilot}}}}\beta _k^2}}{{{\tau ^{\textit{ul}}}K{p_{{\textit{pilot}}}}{\beta _k} + {\sigma ^2}}}, k=1,2,...,K.
\end{equation}
Moreover, $\bm{\varepsilon}$ is independent of $\textbf{G}$ and $\hat{\textbf{G}}$. Now, we can obtain the following lower bounds on the achievable uplink data rate of the $k$th user under imperfect CSI \cite{Ngo2013}:
\begin{equation}\label{equ38}
  \widetilde {R}_{k}^{\textit{MRC,ul}}=\log_2{\left({1+\frac{(M-1)\hat{\beta}_k}{\sum_{i=1,i\neq k}^{K}\hat{\beta}_i+\sum_{i=1}^{K}\gamma_i+\frac{\sigma^2}{p^{\textit{ul}}}}}\right)}
\end{equation}
\begin{equation}\label{equ39}
  \widetilde {R}_{k}^{\textit{ZF,ul}}=\log_2{\left({1+\frac{(M-K)\hat{\beta}_k}{\sum_{i=1}^K \gamma_i+\frac{\sigma^2}{p^{\textit{ul}}}}}\right)}
\end{equation}
where the constraints $M\geq K$ and $M\geq K+1$ are required for (\ref{equ38}) and (\ref{equ39}), respectively. For the downlink case, the vector of received signals can be expressed as
\begin{IEEEeqnarray}{rCl}\label{equ40}
  \textbf{y}^{\textit{dl}}&=&\sqrt{p^{\textit{dl}}}\textbf{G}^H\hat{\textbf{W}}\textbf{x}+\textbf{n} \nonumber\\
  &=&\sqrt{p^{\textit{dl}}}\hat{\textbf{G}}^H\hat{\textbf{W}}\textbf{x}-\sqrt{p^{\textit{dl}}}\bm{\varepsilon}^H\hat{\textbf{W}}\textbf{x}+\textbf{n}.
\end{IEEEeqnarray}
The received signal at the $k$th user is
\begin{IEEEeqnarray}{rCl}\label{equ41}
  \hat{y}_k^{\textit{dl}}=\sqrt{p^{\textit{dl}}}\hat{\textbf{g}}_k^H \hat{\textbf{w}}_k x_k&+&\sqrt{p^{\textit{dl}}}\sum_{i=1,i\neq k}^{K}\hat{\textbf{g}}_k^H \hat{\textbf{w}}_i x_i \nonumber\\
  &-&\sqrt{p^{\textit{dl}}}\sum_{i=1}^{K}\varepsilon_k^H \hat{\textbf{w}}_ix_i+\textbf{n}.
\end{IEEEeqnarray}
Then, we can obtain the following lower bounds on the achievable downlink data rate to the $k$th user under imperfect CSI \cite{Ngo2013}:
\begin{equation}\label{equ42}
  \widetilde {R}_{k}^{\textit{MRT,dl}}=\log_2{\left({1+\frac{(M-1)\hat{\beta}_k}{(K-1)\hat{\beta}_k+K\gamma_k+\frac{\sigma^2}{p^{\textit{dl}}}}}\right)}
\end{equation}
\begin{equation}\label{equ43}
  \widetilde {R}_{k}^{\textit{ZF,dl}}=\log_2{\left({1+\frac{(M-K)\hat{\beta}_k}{K\gamma_k+\frac{\sigma^2}{p^{\textit{dl}}}}}\right)}
\end{equation}
Similarly, $M\geq K$ and $M\geq K+1$, respectively, are required to be satisfied for (\ref{equ42}) and (\ref{equ43}).

\section{Energy Efficiency under a Constraint on the Number of RF Chains}
We have obtained lower bounds on the achievable data rate with different linear processing methods under perfect or imperfect CSI in the previous section. In this section, we formulate the EE of massive MIMO systems with a limitation on the number of RF chains. First, we introduce the power consumption model for massive MIMO systems. Subsequently, we express the energy-efficiency maximization problem in massive MIMO systems subject to a constraint on the number of RF chains.
\subsection{Power Consumption Model}
We use the power consumption model proposed in \cite{Bjornson2015}. For the sake of completeness in the paper, we describe the main characterizations here. The total power consumption consists of the power consumed by the power amplifiers, transceiver chains, channel estimation process, channel coding and decoding units, load-dependent backhaul, linear processing at the BS, and the fixed power consumption (such as power required for site-cooling, control signaling, load-independent power of backhaul infrastructure and baseband processors). Now, the total power can be expressed as follows:
\begin{equation}\label{equ19}
  P_{\textit{sum}}=c+dM+fR
\end{equation}
where
\begin{IEEEeqnarray}{rCl}\label{equ223}
c=
\begin{cases}
   c_1 &\text{MRC/MRT}, \\
   c_1+\frac{BK^3}{3UL_{\textit{BS}}} &\text{ZF/MRT, MRC/ZF and ZF/ZF}, \\
\end{cases}
\end{IEEEeqnarray}
\begin{IEEEeqnarray}{rCl}\label{equ224}
  d=
  \begin{cases}
  d_1+\frac{3BK}{UL_{\textit{BS}}} &\text{MRC/MRT}, \\
  d_1+\frac{B(3K^2+3K+K)}{UL_{\textit{BS}}} &\text{ZF/MRT and MRC/ZF}, \\
  d_1+\frac{B(3K^2+K)}{UL_{\textit{BS}}} &\text{ZF/ZF}, \\
  \end{cases}
\end{IEEEeqnarray}
\begin{equation}\label{equ22}
  f=P_{\textit{COD}}+P_{\textit{DEC}}+P_{\textit{BT}}, {\kern 85pt}
\end{equation}
\begin{IEEEeqnarray}{rCl}\label{equ222}
{c_1} =&& K\left( {\frac{{{p^{\textit{ul}}}}}{{{\eta ^{\textit{ul}}}}} + \frac{{{p^{\textit{dl}}}}}{{{\eta ^{\textit{dl}}}}}} \right) + \frac{{2B{K^2}}}{U}\left( {\frac{{{\tau ^{\textit{ul}}}M}}{{{L_{\textit{BS}}}}} + \frac{{2{\tau ^{\textit{dl}}}}}{{{L_U}}}} \right) \nonumber\\
 &&{\kern 70pt}+ {P_{\textit{FIX}}} + {P_{\textit{SYN}}} + K{P_U},
\end{IEEEeqnarray}
\begin{IEEEeqnarray}{rCl}\label{equ225}
  {d_1} = {P_{\textit{BS}}} + \frac{{2BK}}{{{L_{\textit{BS}}}}}\left( {1 - \frac{{({\tau ^{\textit{ul}}} + {\tau ^{\textit{dl}}})K}}{U}} \right). {\kern 60pt}
\end{IEEEeqnarray}
Above, different expressions in (\ref{equ223}) and (\ref{equ224}) are for different combinations of uplink and downlink linear processing schemes at the BS. For instance, MRC/MRT notation describes that MRC is used at the BS for reception in uplink and MRT is employed at the BS for downlink transmission. The differences in (\ref{equ223}) and (\ref{equ224}) are the results of differences in the power consumption of different linear processing methods. The description of the parameters in these equations along with their typical values are given in Table \ref{tab1}. Readers can also refer to \cite{Bjornson2015} for more details.

\begin{table}[!h]
\footnotesize
\renewcommand{\arraystretch}{1.3}
\caption{Baseband power consumption parameters}
\label{tab1}
\centering
\begin{tabular}{|c|c|c|}
\hline
\bfseries Parameter description & \bfseries Value \\
\hline
Power amplifier efficiency at the users: $\eta^{\textit{ul}}$  & 0.3 \\
Power amplifier efficiency at the BS: $\eta^{\textit{dl}}$ & 0.39 \\
Fixed power consumption: $P_{\textit{FIX}}$ & 18 W \\
Power consumed by local oscillator at BS: $P_{\textit{SYN}}$ & 2 W \\
Power for the circuit components at each user: $P_{U}$ & 0.1 W \\
Power for the circuit components at BS: $P_{\textit{BS}}$ & 1 W \\
Computational efficiency at BS: $L_{\textit{BS}}$ & 12.8 Gflops/W \\
Power required for coding of data signals: $P_{\textit{COD}}$ & 0.1 W/(Gbit/s) \\
Power required for decoding of data signals: $P_{\textit{DEC}}$ & 0.8 W/(Gbit/s) \\
Power required for backhaul traffic: $P_{\textit{BT}}$ & 0.25 W/(Gbit/s) \\
\hline
\end{tabular}
\end{table}
\subsection{Energy Efficiency for ZF/ZF under RF Chain Constraints}
As the formula of the lower bounds on the achievable data rate with different linear processing methods are similar, we consider only ZF receiver in uplink and ZF precoding in downlink in this subsection. Note that since the formulas for the lower bounds on the achievable data rate under imperfect CSI are almost the same as those under perfect CSI, we only formulate the energy-efficiency maximization problem under perfect CSI. The results for the combination of other linear processing methods under perfect or imperfect CSI can be obtained easily following the same procedure introduced in this paper. For the sake of brevity in notations, we will eliminate the subscript ``ZF/ZF'' in the following equations for the results of ZF receiver in uplink and ZF precoding in downlink. The lower bound on the EE with ZF reception in uplink and ZF precoding in downlink under perfect CSI can be expressed as

\begin{IEEEeqnarray}{rCl}\label{equ23}
&&EE({\cal M},{\cal K})
=\frac{\widetilde {R}({\cal M},{\cal K})}{c+dM+f\widetilde {R}({\cal M},{\cal K})}
\end{IEEEeqnarray}
where
\begin{equation}\label{equ500}
  \widetilde {R}({\cal M},{\cal K})=a\sum_{k=1}^{K}\widetilde {R}_{k}^{\textit{ZF,ul}}+b\sum_{k=1}^{K}\widetilde {R}_{k}^{\textit{ZF,dl}}.
\end{equation}
Our goal is to find a subset of antennas ${\cal M}_1$ at the BS and a subset of users ${\cal K}_1$ which maximize $EE(\cal M,\cal K)$ under a limitation on the number of RF chains. Therefore, the original energy-efficiency maximization problem could be written as follows:
\begin{IEEEeqnarray}{rCl}\label{equ24}
  \underset{{\cal M}_1\subseteq{\cal M},{\cal K}_1\subseteq{\cal K}}{\text{maximize}}{\kern 5pt}&& EE({\cal M}_1,{\cal K}_1) \nonumber\\
  {\text{subject to}}{\kern 5pt}&&\left| {{{\cal M}_1}} \right| \le {N_{\textit{RF}}},\left| {{{\cal K}_1}} \right| \le \left| {{{\cal M}_1}} \right|-1
 \end{IEEEeqnarray}
where $N_{\textit{RF}}$ is the number of available RF chains at the BS.

\newcounter{tempequationcounter}
\begin{figure*}[!t]
	\normalsize
	\setcounter{tempequationcounter}{\value{equation}}
	\begin{IEEEeqnarray}{rCl}
		\setcounter{equation}{41}
		\phi ({\textbf{x}}) = \frac{{a\sum\limits_{k = 1}^K {{{\log }_2}} \left( {1 + \frac{{{p^{\textit{ul}}}({M_1} - {K_1}){x_k}{\beta _k}}}{{{\sigma ^2}}}} \right) + b\sum\limits_{k = 1}^K {{{\log }_2}} \left( {1 + \frac{{{p^{\textit{dl}}}({M_1} - {K_1}){x_k}{\beta _k}}}{{{\sigma ^2}}}} \right)}}{{c + dM + f\left( {a\sum\limits_{k = 1}^K {{{\log }_2}} \left( {1 + \frac{{{p^{\textit{ul}}}({M_1} - {K_1}){x_k}{\beta _k}}}{{{\sigma ^2}}}} \right) + b\sum\limits_{k = 1}^K {{{\log }_2}} \left( {1 + \frac{{{p^{\textit{dl}}}({M_1} - {K_1}){x_k}{\beta _k}}}{{{\sigma ^2}}}} \right)} \right)}}
	\end{IEEEeqnarray}
	\setcounter{equation}{\value{tempequationcounter}}
	\hrulefill
	\vspace*{4pt}
\end{figure*}

The above problem is an NP-hard problem, and it cannot be solved analytically. We assume $\textbf{x}$ is an $N\times 1$ vector, where $N=K+M$, and $x_i\in \{0,1\}$ for all $i \in \{1,\ldots, N\}$. The first $K$ elements of $\textbf{x}$ correspond to the user selection results, and the latter $M$ elements correspond to antennas. Define $\phi ({\textbf{x}})$ as shown in (41) at the top of next page,
where ${K_1} = \sum\nolimits_{i = 1}^K {{x_i}} $ is the number of selected users, and ${M_1} = \sum\nolimits_{i = K + 1}^N {{x_i}}$ is the number of selected antennas. Then, the optimization problem in (\ref{equ24}) can be rewritten as
\addtocounter{equation}{1}
\begin{IEEEeqnarray}{rCl}\label{equ35}
  \text{maximize}&&{\kern 5pt}\phi(\textbf{x}) \nonumber\\
  \text{subject to}&&{\kern 5pt}\sum\limits_{i = K + 1}^N {{x_i}} \leqslant {N_{\textit{RF}}}, \nonumber\\
                   &&{\kern 5pt}\sum\limits_{i = 1}^K {{x_i}}  \leqslant \sum\limits_{i = K + 1}^N {{x_i}}  - 1, \nonumber\\
                   &&{\kern 5pt}x_i\in\{0,1\}, i=1,2,...,K,
\end{IEEEeqnarray}
which is a typical combinatorial optimization problem, and it can be solved efficiently via the Gibbs-sampling based method.
In the following sections, we study how to obtain the energy-efficiency maximizing subset of users and antennas via a learning-based stochastic gradient descent method.

\section{Learning-Based Stochastic Gradient Descent Combinatorial Optimization Algorithm}
In this section, we analyze how to solve the combinatorial optimization problem using the learning-based stochastic gradient descent method. For the completeness of the paper, we initially review the original learning-based stochastic gradient descent method proposed in \cite{Berny2001}, and then address its advantages and disadvantages in solving the problem in (\ref{equ35}). Following this, we provide several extensions to the original learning-based stochastic gradient descent method to make it an efficient algorithm for solving the problem in (\ref{equ35}).
\subsection{Learning-Based Stochastic Gradient Descent Method}
Suppose we have a combinatorial optimization problem with $N$ features ${\textbf{x}} = {[{x_1},{x_2},...,{x_N}]^T}$, ${x_i} \in \{ 0,1\} $, and we want to minimize the objective function $\varphi ({\textbf{x}})$. In \cite{Berny2001}, a learning-based stochastic gradient descent algorithm is proposed for solving this problem based on the characteristics of Gibbs distribution and dynamical systems.

The Gibbs distribution maps each value of the objective function onto a probability defined by
\begin{equation}\label{equg1}
 p_T^ * ({\textbf{x}}) = \frac {\exp ( - \varphi({\textbf{x}})/T)} {\sum\nolimits_{{\textbf{y}} \in {\cal S}} {\exp ( - \varphi({\textbf{x}})/T)} }
\end{equation}
where $T>0$ is the analogue of a temperature and ${\cal S} = {\{ 0,1\} ^N}$ is the set of all possible ${\textbf{x}}$. Let us define ${\cal U} = \{ {\textbf{x}} \in {\cal S}|\forall {\textbf{y}} \in {\cal S},\varphi({\textbf{x}}) \le \varphi({\textbf{y}})\} $. The Gibbs distribution $p_T^ * $ converges to a uniform distribution on ${\cal U}$ when $T$ tends to zero. In other words, we can obtain the optimal solutions by finding the limit distribution of $p_T^ * $ as $T \to 0$. However, this is difficult in practice. So instead of finding the limit distribution directly, we search for a distribution which has the smallest Kullback-Leibler (KL) divergence to an implicit Gibbs distribution. The KL divergence between $p$ and $p_T^ * $ is
\begin{equation}\label{equg2}
  D(p,p_T^ * ) =  - \sum\limits_{{\textbf{x}} \in {\cal S}} {p({\textbf{x}}) \ln{\frac {p_T^ * ({\textbf{x}})}  {p({\textbf{x}})}}}
\end{equation}
where $\ln( \cdot )$ is the natural logarithm function. Then, the problem is converted into the minimization of the free energy of the system:
\begin{equation}\label{equg3}
  F = \sum\nolimits_{{\textbf{x}} \in {\cal S}} {p({\textbf{x}})\varphi({\textbf{x}}) + T\sum\nolimits_{{\textbf{x}} \in {\cal S}} {p({\textbf{x}}){\rm{ln}}(p({\textbf{x}}))} } .
\end{equation}

This is still a discrete problem, and it is not easy to solve in practice. By introducing ${\bm{\theta }} = {[{\theta _1},{\theta _2},...,{\theta _N}]^T}$ as the probability distribution parameter for the $N$-dimensional random vector ${\textbf{x}}={[{x _1},{x _2},...,{x _N}]^T}$, we can convert
the discrete optimization into a continuous optimization problem. Then, we introduce the following dynamical system
\begin{equation}\label{equg4}
  {\frac {\partial \theta } {\partial t}} + {\frac {\partial F} {\partial \theta }} = 0.
\end{equation}
At last, we obtain the following statistical update rule:
\begin{equation}\label{equ45}
{\theta _i}(t + 1) = {\theta _i}(t) - \alpha \left(\varphi({\textbf{x}}) + T(1 + \ln(p({\textbf{x}})))\right){\frac {\partial \ln(p({\textbf{x}}))} {\partial {\theta _i}}}
\end{equation}
where $\alpha$ is the learning rate, and ${\frac {\partial \ln(p({\textbf{x}}))} {\partial {\theta _i}}}$ is the gradient.

Suppose we choose the random variables $x_i$ to be independent binomially distributed. Then, the joint distribution of the random vector $\textbf{x}$ is as follows:
\begin{equation}\label{equ46}
  p({\textbf{x}}) = \prod\limits_{i = 1}^N {({x_i}{p _i} + (1 - {x_i})(1 - {p_i}))}
\end{equation}
where $p_i$ is the probability that $x_i$ equals to 1. The relationship between $p_i$ and $\theta_i$ is
\begin{equation}\label{equ47}
  {p_i} = \frac{1}{2}(1 + \tanh (\beta {\theta _i})),
\end{equation}
and the gradient of $\ln(p({\textbf{x}}))$ is
\begin{equation}\label{equ48}
  {\frac {\partial \ln(p({\textbf{x}}))} {\partial {\theta _i}}} = 2\beta ({x_i} - {p_i}).
\end{equation}
With this, the update rule in (\ref{equ45}) becomes
\begin{equation}\label{equup}
  {\theta _i}(t + 1) = {\theta _i}(t) - 2\alpha \beta \left( {\varphi({\textbf{x}}) + T(1 + \ln(p({\textbf{x}})))} \right)({x_i} - {p_i}).
\end{equation}

As noted before, the Gibbs distribution converges to a uniform distribution which achieves the optimal solutions, and Algorithm 1 below will obtain the optimal solutions with sufficiently many iterations. If we use the Metropolis algorithm to update the Gibbs distribution to get the optimal solutions, then we have the well-known simulated annealing algorithm. As the objective function decreases fastest in the gradient direction, Algorithm 1 will arrive at the optimal solution quicker than the simulated annealing algorithm.

If we are interested in finding the maximum value of an objective function, we just need to add a negative sign before the objective function, and then substitute it into our algorithm.

Below, we provide the learning-based combinatorial optimization algorithm using the stochastic gradient descent as Algorithm 1.
\begin{algorithm}
 \caption{Learning-based combinatorial optimization algorithm.}
 \begin{algorithmic}[1]
  \STATE Initialize the multivariate Bernoulli distribution parameter $\bm{\theta}$, learning rate $\alpha$, and $\beta$;
  \STATE Generate an $N$-dimensional Bernoulli random vector $\textbf{x}$ which satisfies all the constraints in the combinatorial optimization problem, with parameter vector $\bm{\theta}$;
  \STATE Calculate the objective function value $\varphi(\textbf{x})$;
  \STATE Update $\bm{\theta}$ with (\ref{equup});
  \STATE If the stop criteria is met, stop; otherwise, go to step 2.
 \end{algorithmic}
\end{algorithm}
\subsection{Drawbacks of Algorithm 1 and the Corresponding Solutions}
Although Algorithm 1 could arrive at the maximum value of the objective function, fluctuations occur, slowing the convergence.

Another drawback lies in step 2 of Algorithm 1, which should generate an $N$-dimensional multivariate Bernoulli random vector $\textbf{x}$ which satisfies all the constraints with parameter vector $\bm{\theta}$. It may become difficult to produce this $N$-dimensional multivariate Bernoulli random vector directly as the constraints for the combinatorial optimization problem are strict, e.g., especially when the probability of the $N$-dimensional multivariate Bernoulli random vector is very small, such as less than ${10^{ - 10}}$. Such low probabilities may be experienced in practice and such cases require a large number of samples to get a realization of the event.

These observations motivate us to provide the following modifications to Algorithm 1 and use an efficient rare event simulation method to overcome these drawbacks:
\subsubsection{Convergence}
 The fluctuation problem of Algorithm 1 could be solved by some revisions in steps 2 and 3. Instead of generating only one sample, we generate a population of $N_1$ individuals from the $N$-dimensional multivariate Bernoulli distribution, which satisfy the constraints with parameter vector $\bm{\theta}$ in step 2. Then we determine the objective function value for each individual and select the one which leads to the smallest value in step 3. Numerical results demonstrate that Algorithm 1 converges to the optimal solutions quickly with these changes.
\subsubsection{Rare event simulation}
The second drawback of Algorithm 1 is overcome with the efficient subset simulation method for rare event estimation, proposed by Au and Beck in \cite{AuBeck2001}. The basic idea of subset simulation method is to decompose the rare event $F$ into a sequence of progressively ``less-rare" nested events $F = {F_m} \subset {F_{m - 1}} \subset  \cdots  \subset {F_1}$, where ${F_1}$ is a relatively frequent event \cite{Zuev2021}. Then the small probability $P(F)$ of the rare event $F$ can be represented as
\begin{equation}\label{equ50}
  P(F)=P(F_1)\cdot P(F_2|F_1)\cdot ... \cdot P(F_m|F_{m-1})
\end{equation}
where $P(F_k|F_{k-1})=P(F_k)/P(F_{k-1})$ is the conditional probability of $F_k$ given the occurrence of $F_{k-1}$, for $k=2,...,m$. With this, the estimation of the rare event problem is transferred to the product of relatively frequent events. In practice, it is always not obvious how to decompose the rare event into a sequence of relatively frequent events. This could be done adaptively via the Markov chain Monte Carlo technique \cite{Zuev2021}.

With the former two revisions, we obtain the learning-based stochastic gradient descent algorithm for EE maximization as described in Algorithm \ref{alg1} below.
\begin{algorithm}
 \caption{Learning-based stochastic gradient descent algorithm for EE maximization.}
 \label{alg1}
 \begin{algorithmic}[1]
  \STATE Initialize the multivariate Bernoulli distribution parameter ${\bm{\theta }}$, learning rate $\alpha$, $\beta$, and the number of samples at each iteration $N_1$;
  \STATE Generate $N_1$ $N$-dimensional multivariate Bernoulli random vectors $\textbf{x}$ with parameter vector $\bm{\theta}$;
  \STATE If none of the $N_1$ multivariate Bernoulli random vectors $\textbf{x}$ generated in step 2 satisfies the constraints in (\ref{equ35}), we use the subset simulation method introduced in this section to produce another $N_1$ multivariate Bernoulli random vectors $\textbf{x}$, instead of the vectors produced in step 2.
  \STATE Calculate $\phi ({\bf{x}})$ for each of the multivariate Bernoulli random vectors $\textbf{x}$ generated in step 2 or 3, which satisfy the constraints in (\ref{equ35}). Then, select the one that maximize $\phi ({\bf{x}})$;
  \STATE Update $\bm{\theta}$ with (\ref{equup}). As the goal is to maximize the EE, we substitute $\varphi({\textbf{x}}) = -\phi ({\bf{x}})$ into (\ref{equup});
  \STATE If the stop criteria is met, stop; otherwise, go to step 2.
 \end{algorithmic}
\end{algorithm}

We usually stop learning when EE converges to some value. The subset of antennas at the BS and subset of users are selected jointly in Algorithm 2. Numerical simulation results in the following section will show that Algorithm 2 is very efficient to solve the EE maximization problem subject to a limitation on the number of RF chains.

In order to see the effect of RF chains constraint on the maximum EE that the system could achieve, We also consider the case where there is no RF chains constraint for the selected number of antennas. In this situation, it is equivalent to the case that the number of RF chains constraint equals the total number of antennas at the BS, i.e., ${N_{\textit{RF}}} = M$ in equation (\ref{equ35}).

\section{Numerical Results}

In this section, we provide numerical results to analyze the performance. More specifically, we primarily focus the EE achieved when a ZF receiver is used at the BS for uplink data reception, and ZF precoding is employed at the BS for downlink data transmission (abbreviated as the ZF/ZF strategy). We analyze the EE as a function of the SNR and the number of available RF chains. Performance levels achieved with other uplink receivers and downlink precoders, addressed in Section \ref{sec:linearprocessing}, can be determined similarly. In Section \ref{subsec:comparison}, we compare our learning-based stochastic gradient descent method with the algorithm proposed in \cite{Bjornson2015}, and demonstrate that our algorithm can further improve the EE of the system.

In Section \ref{subsec:converge}, \ref{subsec:complexity} and \ref{subsec:ZF/ZF}, we consider a single-cell massive MIMO system with a radius of 1000 m. Users are randomly distributed in the cell, and we assume that there is no user within the radius of 100 m. The path loss decay exponent is $\gamma=3.8$, and the shadow fading has a standard deviation of $\sigma_{\textit{shad}}=8$ dB. The transmission bandwidth is $B=20$ MHz, while the channel coherence bandwidth is $B_c=180$ kHz. The number of symbols transmitted during a time-frequency coherence block is $U=1800$. The total noise power at the BS is $-96$ dBm. The relative pilot length during uplink and downlink channel estimation is ${\tau ^{\textit{ul}}} = {\tau ^{\textit{dl}}} = 1$, i.e., the number of pilots equal to the number of users. The uplink and downlink transmission ratios are ${\varsigma ^{\textit{ul}}}=0.4$ and ${\varsigma ^{\textit{dl}}}=0.6$. The parameters related to the baseband power consumption model are given in Table \ref{tab1}. The energy efficiency is averaged with 5000 independent realizations.

\subsection{Convergence of Algorithm 2} \label{subsec:converge}
\begin{figure}[htbp]
	\center
	\includegraphics[width=3.3in]{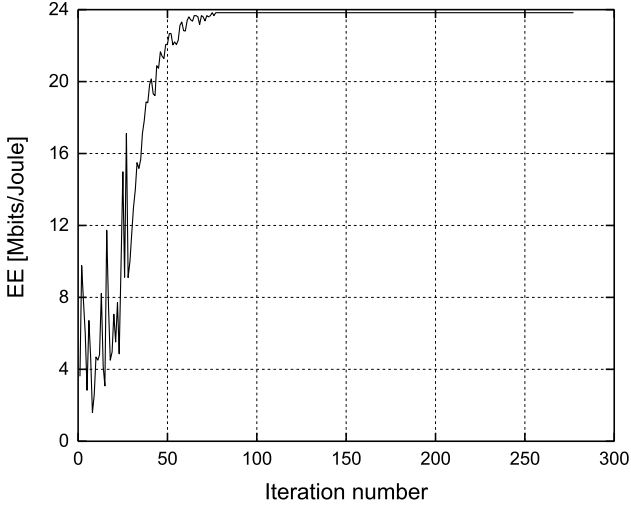}\\
	\caption{EE versus number of iterations under imperfect CSI, SNR=20 dB, RF=15, 60 antennas at the BS and 50 users uniformly distributed in the cell.}\label{figiterone}
\end{figure}

Algorithm 2 can be regarded as a variant of stochastic gradient decent algorithms, which have been widely used to solve the optimization problems with nonconvex objective functions. Indeed, while being different, Algorithm 2 has certain similarities to the stochastic gradient descent algorithms addressed in \cite{Berny2000,Gallagher2005,Zhang2019}, where the convergence properties of such algorithms are discussed in detail. Additionally, while establishing theoretical guarantees for the convergence is challenging, the convergence of Algorithm 2 can be addressed and demonstrated by numerical analysis in the settings considered in the simulations. Fig. \ref{figiterone} plots the achieved EE with Algorithm 2 versus the number of iterations with imperfect CSI under RF chain constraints for a single realization of the channel coefficients. We observe that Algorithm 2 converges after about 70 iterations. Numerical results show that the convergence tendency of Algorithm 2 under different settings are similar to Fig. \ref{figiterone}.

Moreover, since the optimal solution of combinatorial optimization problems can be obtained by exhaustive search, it is obvious that if it achieves the same performance as exhaustive search, Algorithm 2 converges to the optimal solution. In order to perform exhaustive search, we consider a relatively low-dimensional setting and assume that there are 8 antennas at the BS, and 6 users uniformly distributed in the cell.

\begin{figure}[htbp]
	\center
	\includegraphics[width=3.3in]{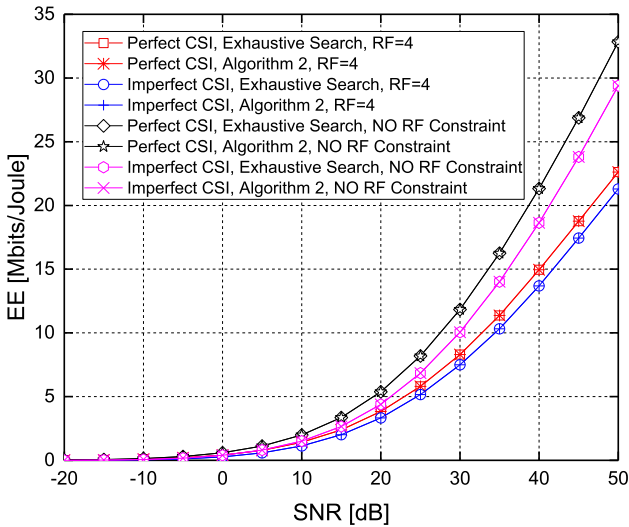}\\
	\caption{Comparison of the maximum EE levels versus SNR achieved with Algorithm 2 and exhaustive search method in single-cell massive MIMO systems.}\label{figconv}
\end{figure}

In Fig. \ref{figconv}, we compare the maximum EE levels achieved with Algorithm 2 and also with exhaustive search in single-cell massive MIMO systems, assuming both perfect and imperfect CSI, with and without a limitation on the number of available RF chains. We observe that the performances of Algorithm 2 and exhaustive search are indistinguishable, and hence Algorithm 2 can attain approximately optimal solutions for problem of joint antenna selection and user scheduling to maximize the achievable joint uplink and downlink EE, i.e., Algorithm 2 converges to the approximately optimal solution points of the corresponding combinatorial optimization problem.

\begin{figure}[htbp]
	\center
	\includegraphics[width=3.3in]{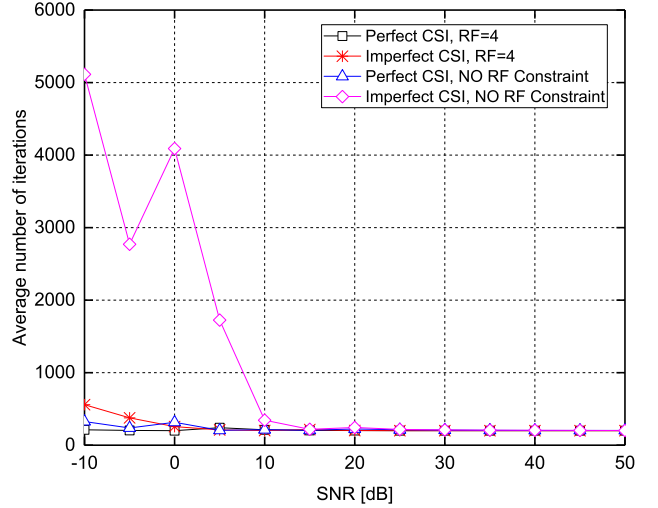}\\
	\caption{Average number of iterations needed for the converge of Algorithm 2.}\label{figiteravg}
\end{figure}

Fig. \ref{figiteravg} plots the average number of iterations (averaged over channel fading) needed for the converge of Algorithm 2. In the numerical simulations, when the successive number of objective function values, whose difference between the former and latter is smaller than a given tolerance, is larger than 200, we deem Algorithm 2 as converged and stop the iteration process. Therefore, the lower bound of the average iteration number is 200. With this stopping criterion, the total number of iterations before Algorithm 2 is stopped in the simulation provided in Fig. \ref{figiterone} is 277 (although convergence is attained after about 70 iterations as seen in the figure).

\subsection{Computational Complexity of Algorithm 2} \label{subsec:complexity}
Since Algorithm 2 is a Gibbs distribution based stochastic gradient descent method, it is difficult to analyze the computational complexity theoretically. Thus, the computational complexity is analyzed via numerical results. From the convergence analysis before, we know that Algorithm 2 can achieve approximately optimal solutions for the corresponding combinatorial optimization problem considered in this paper, which could also be obtained by exhaustive search. Therefore, we here compare the computational complexities of Algorithm 2 and exhaustive search. The former is comprised of generating multivariate Bernoulli random vectors, checking the RF chain constraint, and computing $\phi ({\bf{x}})$, while the latter consists of only checking the RF chain constraint and computing $\phi ({\bf{x}})$. Numerical results demonstrate that the time used for the computation of $\phi ({\bf{x}})$ dominates the computational complexity of both Algorithm 2 and exhaustive search. Therefore, the number of times $\phi ({\bf{x}})$ is computed during the simulations is used as a criterion for the computational complexity comparison. In this section, we consider 60 antennas at the BS, and 50 users uniformly distributed in the cell.

\begin{figure}[t]
	\centering
	\subfigure{
		\label{comp1} 
		\includegraphics[width=3in]{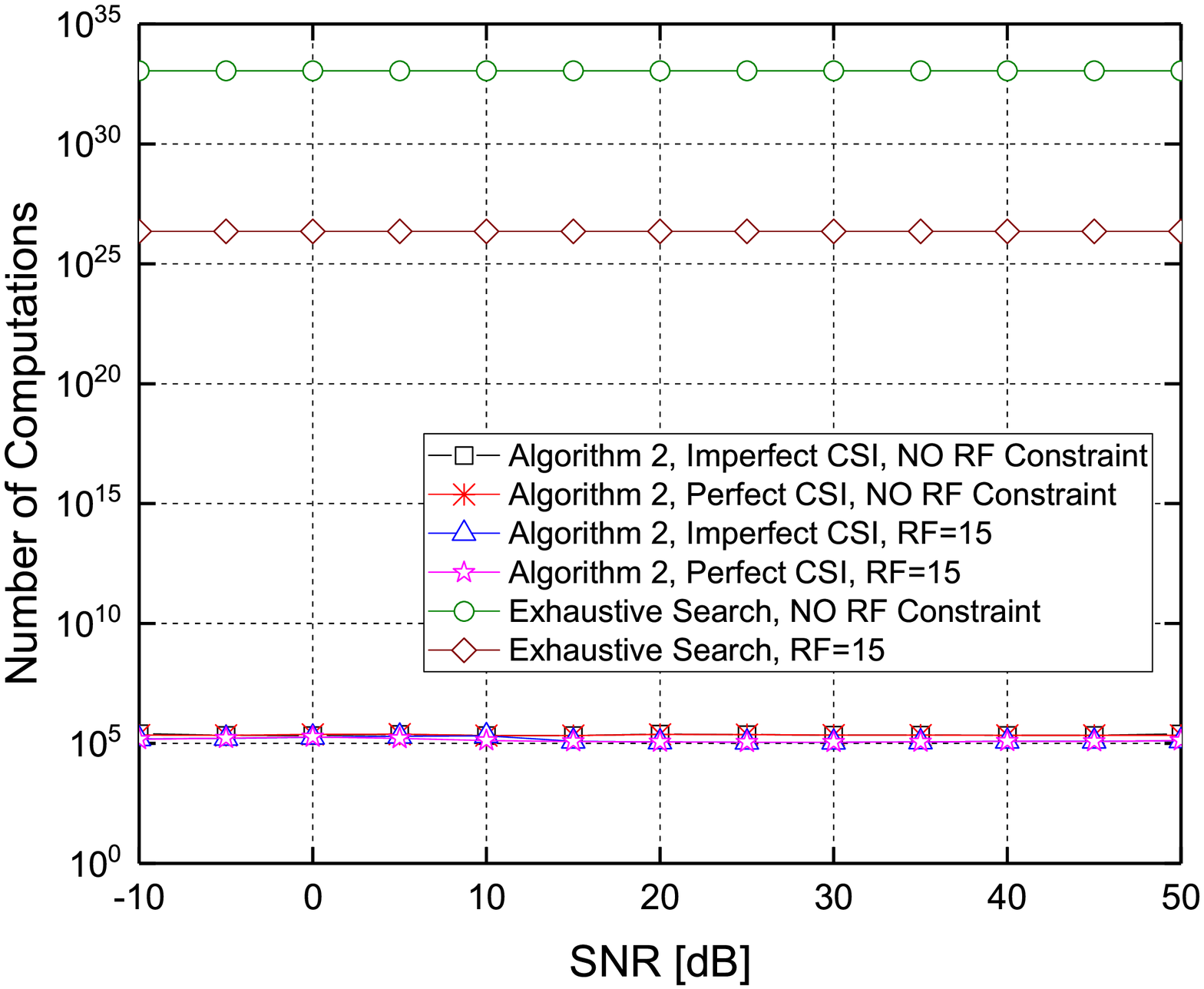}}
	\subfigure{
		\label{comp2} 
		\includegraphics[width=3in]{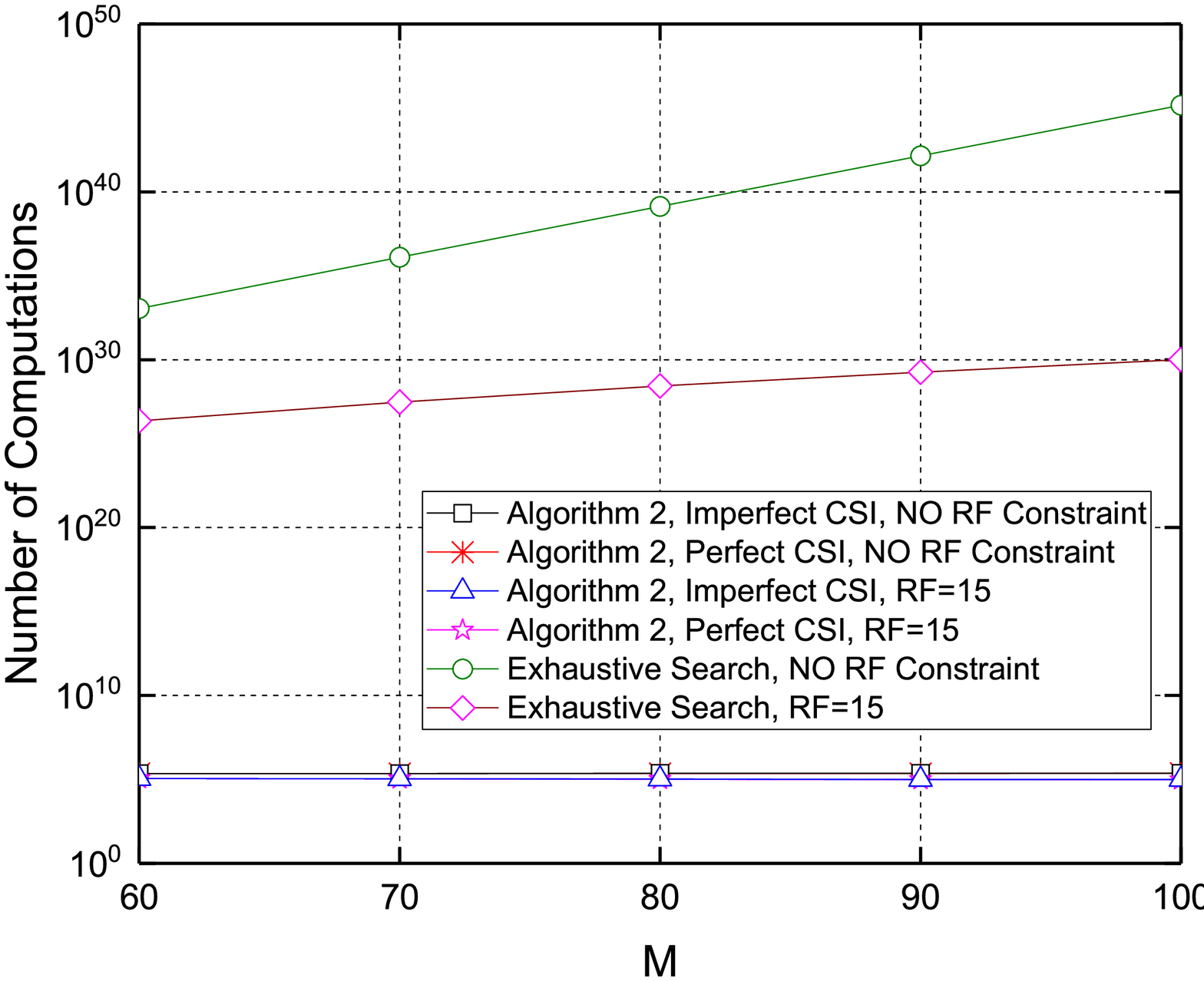}}
	\subfigure{
		\label{comp3} 
		\includegraphics[width=3in]{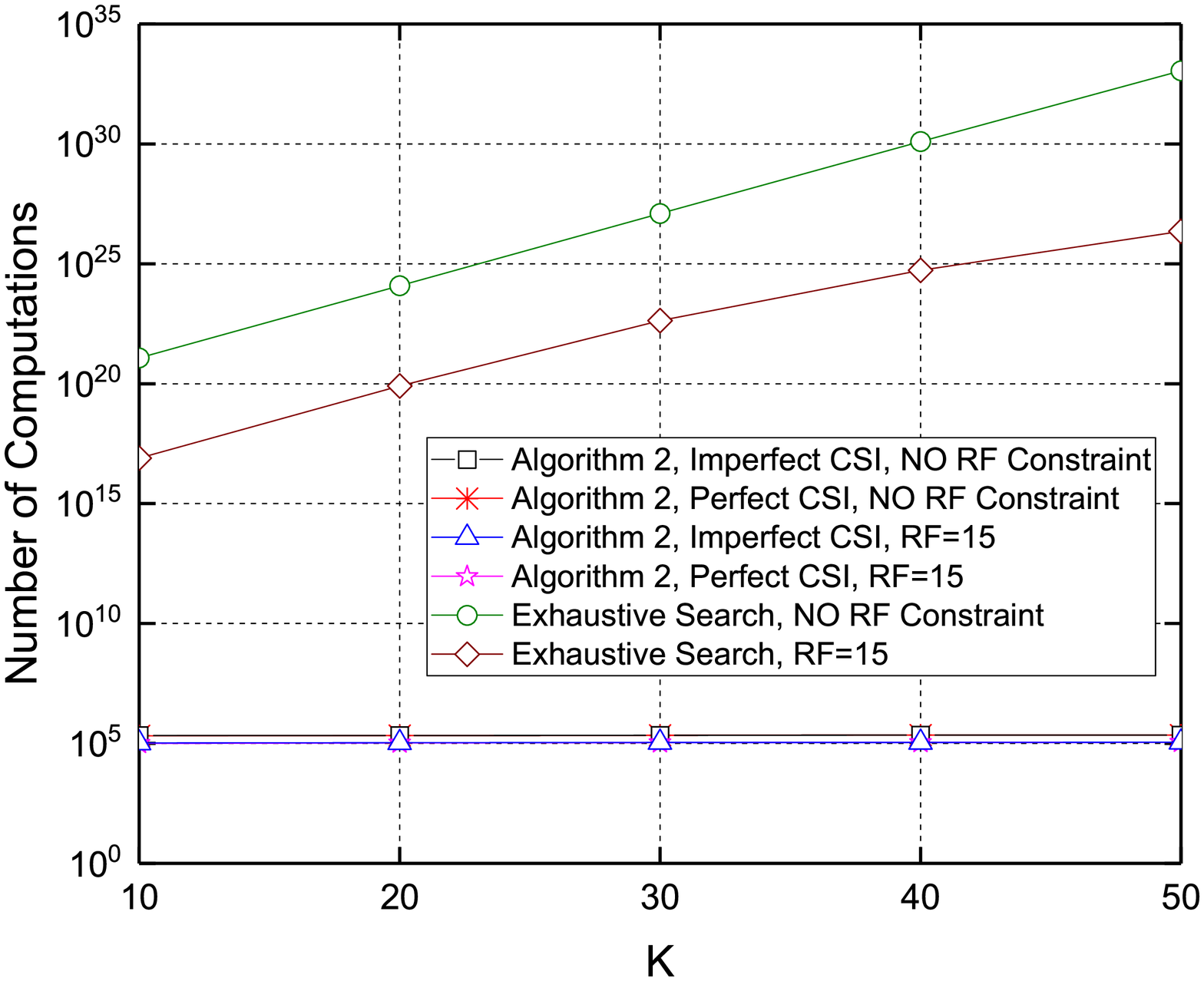}}
	\caption{Average number of computations of $\phi ({\bf{x}})$ vs. SNR, number of BS antennas, $M$, and number of users, $K$, when employing Algorithm 2 and exhaustive search method.}
	\label{comp} 
\end{figure}

Fig. \ref{comp} plots the curves for the computational complexities of Algorithm 2 and the exhaustive search method versus SNR, the number of antennas at BS and the number of users, respectively. The y-axis corresponds to the average number of times $\phi ({\bf{x}})$ is computed during the numerical simulations. All the curves in Fig. \ref{comp} show that the computational complexity of Algorithm 2 is significantly (e.g., many orders of magnitude) less than that of the exhaustive search method.

\subsection{Maximum EE achieved by the ZF/ZF strategy } \label{subsec:ZF/ZF}
\begin{figure}[htbp]
  \center
  \includegraphics[width=3.3in]{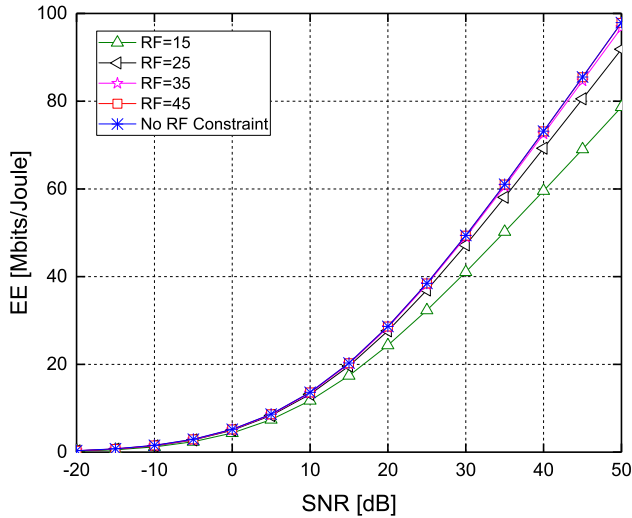}\\
  \caption{Maximum EE achieved in a single-cell massive MIMO system versus SNR with different RF constraints under perfect CSI.}\label{fig5}
\end{figure}

\begin{figure}[htbp]
  \center
  \includegraphics[width=3.3in]{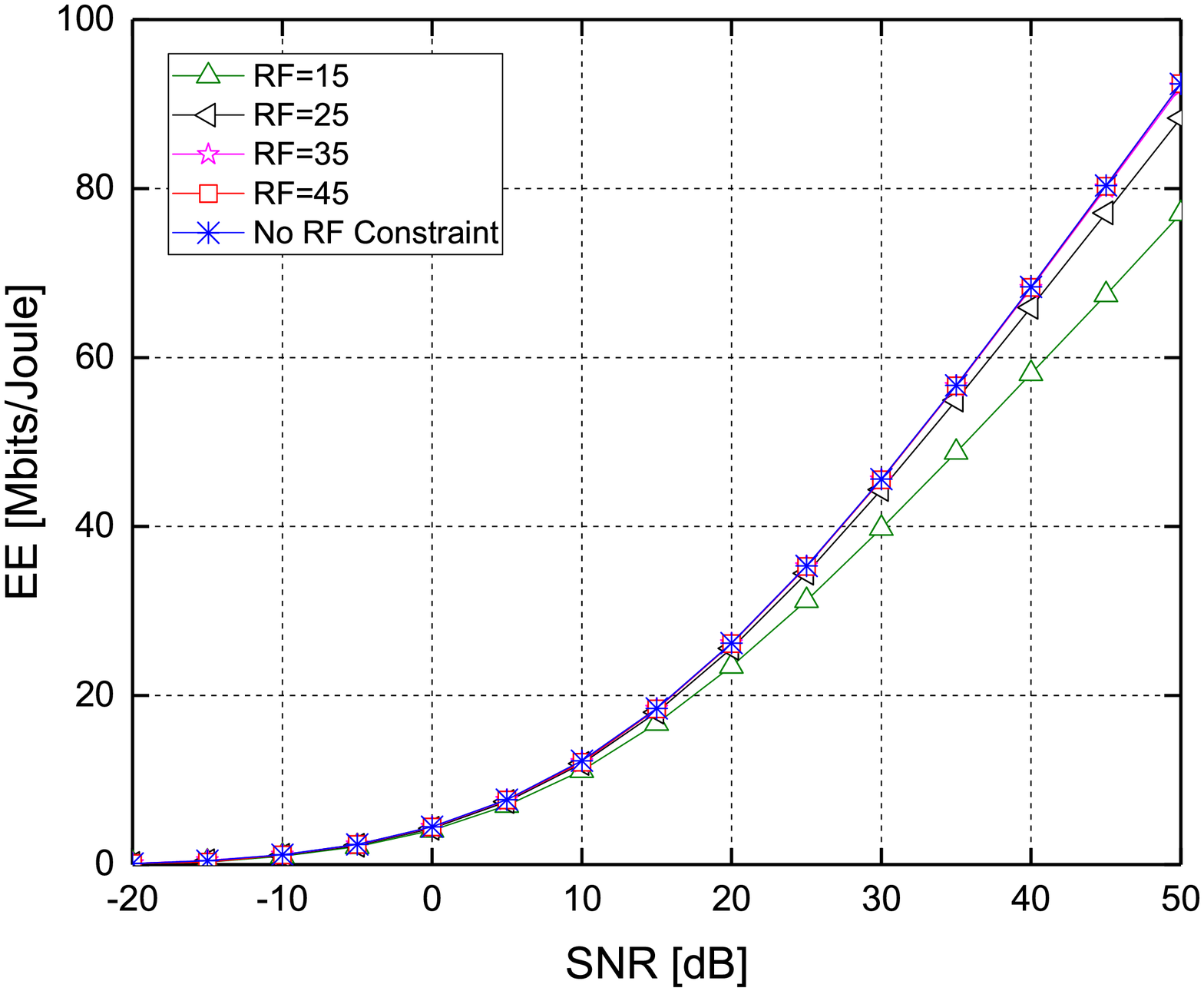}\\
  \caption{Maximum EE achieved in a single-cell massive MIMO system versus SNR with different RF constraints under imperfect CSI.}\label{fig6}
\end{figure}

\begin{figure}[htbp]
  \center
  \includegraphics[width=3.3in]{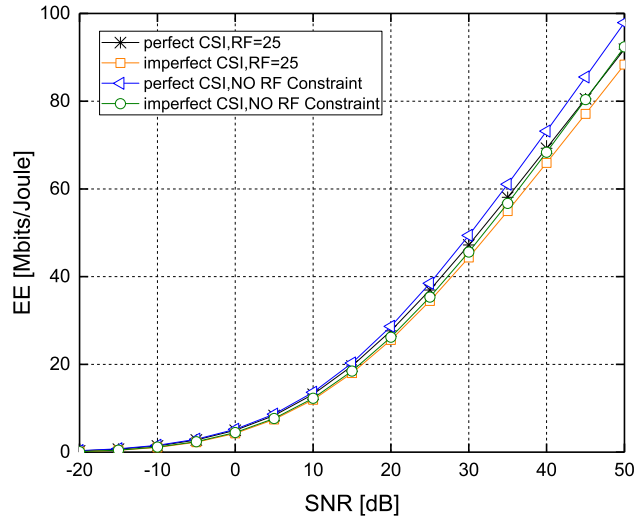}\\
  \caption{Comparison of maximum EE achieved in a single-cell massive MIMO system versus SNR under perfect and imperfect CSI.}\label{fig7}
\end{figure}

In this subsection, we assume that there are 60 antennas at the BS, and 50 users uniformly distributed in the cell. In Figs. \ref{fig5}, \ref{fig6} and \ref{fig7}, we plot the maximum EE (that the single-cell massive MIMO system could achieve) as a function of SNR with perfect and imperfect CSI and with and without RF chain constraints. We observe in Figs. \ref{fig5} and \ref{fig6} that there is almost no difference in EE with and without RF chains constraints at low SNRs, regardless of whether there is perfect (Fig. \ref{fig5}) or imperfect CSI (Fig. \ref{fig6}).
On the other hand, at medium and large SNR levels, the presence of RF chain constraints leads to a noticeable decrease in the maximum EE. We note that the curves approach straight lines in the high SNR regime, but the slopes decrease when SNR is larger than 20 dB as the RF chain constraints become more stringent. Finally, we observe in Fig. \ref{fig7} that compared to perfect CSI case, the maximum EE is smaller under imperfect CSI with the same SNR. However, the curves in this figure demonstrate that the performance gap is less than 10\% under imperfect CSI, compared with that under perfect CSI. Besides, since the magnitude of ${P_{\textit{FIX}}}$, ${P_{\textit{SYN}}}$ and ${P_{\textit{BS}}}$ are very large compared with the summation of uplink and downlink transmitting power, the dominate component of the power consumption model in (\ref{equ19}) increases slower than the increasing rate of system's total achievable data rate. Therefore, other than bell-shaped curves, the maximum EE achieved by the system keeps increasing as the SNR increases.

Additionally, we have the following intriguing observations. Under perfect CSI, we notice in Fig. \ref{fig5} that the EE curve with RF $=45$ overlaps with the one without a constraint on the number of RF chains, while the EE curves with RF $=35$ and RF $=45$ both overlap in Fig. \ref{fig6} with the one without RF chain constraints under imperfect CSI. This indicates that under imperfect CSI, a smaller number of RF chains is needed to attain the same performance level achieved in the absence of RF chain constraints. At the same time, it is important to note that the best performance under imperfect CSI is less than the best performance under perfect CSI.

\begin{figure}[htbp]
  \center
  \includegraphics[width=3.3in]{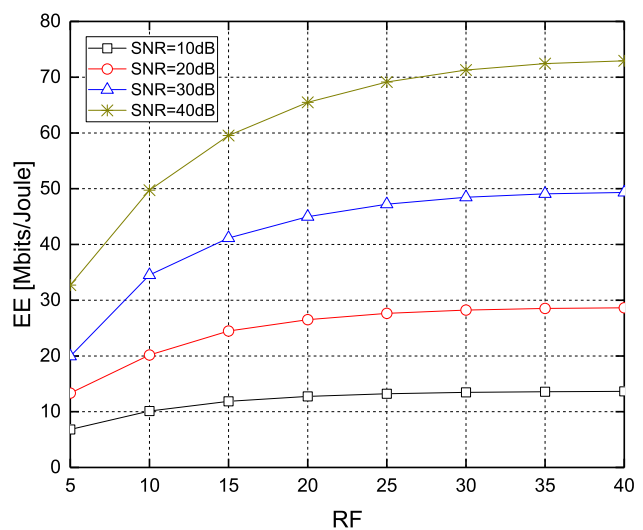}\\
  \caption{Maximum EE achieved in a single-cell massive MIMO system versus RF chains constraint with different SNR under perfect CSI.}\label{fig8}
\end{figure}

\begin{figure}[htbp]
  \center
  \includegraphics[width=3.3in]{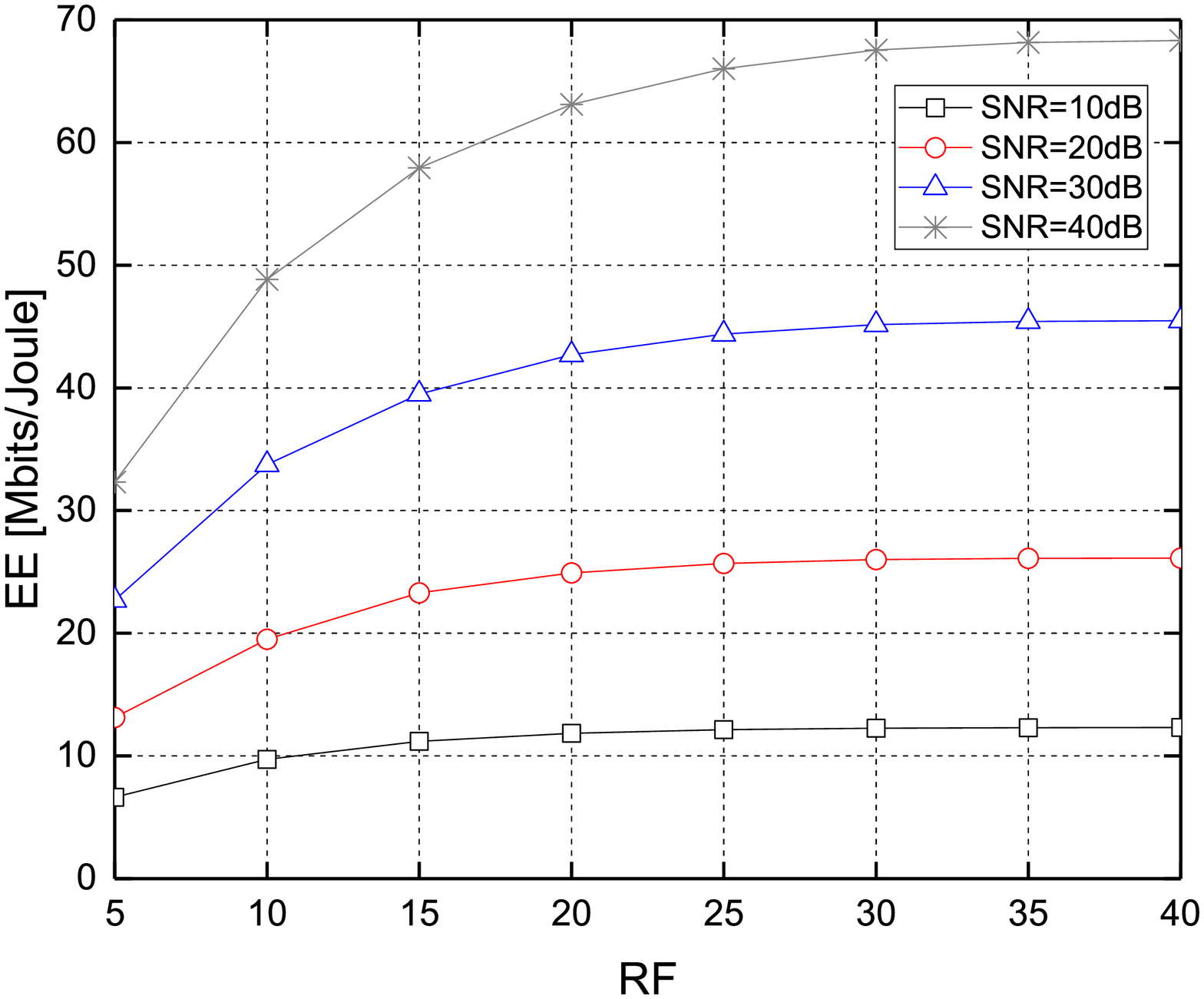}\\
  \caption{Maximum EE achieved in a single-cell massive MIMO system versus RF chains constraint with different SNR under imperfect CSI.}\label{fig81}
\end{figure}

\begin{figure}[htbp]
  \center
  \includegraphics[width=3.3in]{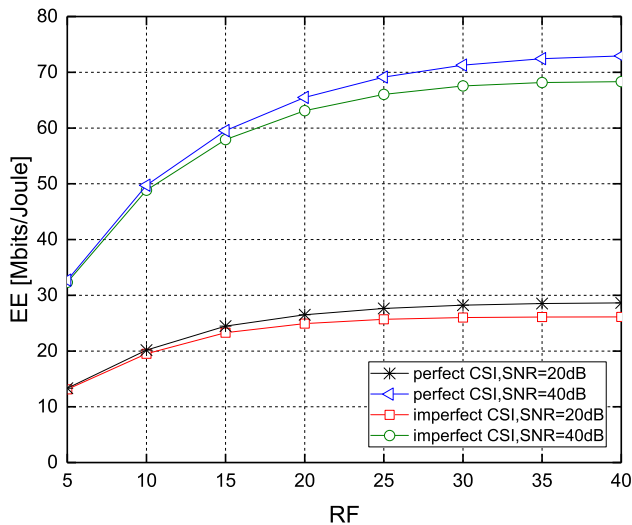}\\
  \caption{Comparison of maximum EE achieved in a single-cell massive MIMO system versus RF chains constraint under perfect and imperfect CSI.}\label{fig51}
\end{figure}

The maximum EE that can be achieved in single-cell massive MIMO systems versus the number of allowed RF chains for different SNRs under perfect and imperfect CSI are plotted in Figs. \ref{fig8}, \ref{fig81} and \ref{fig51}. Under both perfect and imperfect CSI, EE initially increases as the number of allowed RF chains grows, but the rate of increase slows and EE starts staying fixed after a certain threshold on the number of RF chains. We observe that this RF chain threshold is larger at higher SNR levels. However, at the same SNR, the RF thresholds are interestingly equal under perfect and imperfect CSI, although, as noted before, the maximum EE that can be achieved under imperfect CSI is less than that under perfect CSI.

We remark that the results in Fig. \ref{fig81} could guide the design of single-cell massive MIMO systems in practice. As there are large numbers of antennas at the BS, how many RF chains should be employed is a critical consideration in the design of massive MIMO systems.
For instance, in the design of a real system, given the EE requirements, one can determine approximately how many RF chains should be set up at the BS from the results in Fig. \ref{fig81}.
\subsection{Comparison of Algorithm 2 and the Method in \cite{Bjornson2015}} \label{subsec:comparison}
Since \cite{Bjornson2015} also addresses joint uplink and downlink EE maximization in single-cell massive MIMO systems, we in this section compare the performance of Algorithm 2 with that of the method proposed in \cite{Bjornson2015}. Specifically, we assume that the users are distributed in a circular cell with maximum radius ${d_{\textit{max} }} = 250 m$ and minimum radius ${d_{\textit{min} }} = 35 m$. Path-loss is the dominant component in the large-scale fading of users' channels. The large-scale fading is expressed as ${\beta _k} = {\beta _0}/{\left( {{d_k}} \right)^\alpha }$, where $d_k$ is the distance from the $k$th user to the BS, the path-loss decay exponent $\alpha=3.76$, and the large-scale fading at ${d_{\textit{min} }}$ is $\beta_0=35.3$ dB. We assume that there are 220 antennas at the BS, and 150 users uniformly distributed in the cell. The other parameters are the same as described at the beginning of this section.

Algorithm 2 and \cite{Bjornson2015} achieve optimal joint uplink and downlink EE with two entirely different methods. For instance, EE maximization is achieved with joint antenna selection and user scheduling in Algorithm 2, while the algorithm proposed in \cite{Bjornson2015} maximizes EE with random antenna and user selection (Specifically, in \cite{Bjornson2015}, with a specified power allocation algorithm, each user is assumed to attain the same rate. Then, the EE maximization problem reduces to finding the optimal number of antennas and users, which maximize the EE of the system. Given the optimal numbers of antennas and users, the set of users and antennas are selected randomly). In other words, as its most important aspect, Algorithm 2 performs joint antenna selection and user scheduling, while the algorithm proposed in \cite{Bjornson2015} is used to obtain the optimal number of antennas and users. Additionally, Algorithm 2 can be employed under any power allocation scheme, while the algorithm proposed in \cite{Bjornson2015} only works under the power allocation scheme which leads to the same rate for all the users.

In order to have fairness in the comparison, we assume equal total transmit power for both Algorithm 2 and the method in \cite{Bjornson2015}, i.e., the total transmit power for the entire set of users is $B{\varsigma ^{ul}}{\sigma ^2}\rho {S_\beta }K/{\eta ^{ul}}$ or $B{\varsigma ^{dl}}{\sigma ^2}\rho {S_\beta }K/{\eta ^{dl}}$, for uplink or downlink communication, respectively, where the design parameter $\rho  = {p^{\textit{ul}}}{\beta _k}/{\sigma ^2}= {p^{\textit{dl}}}{\beta _k}/{\sigma ^2}$, and ${S_\beta } = {\mathbb E}\{ {\beta ^{ - 1}}\} $. For ZF/ZF strategy, each user's transmit power for the algorithm proposed in \cite{Bjornson2015} is proportional to $1/{\left\| {{{\bf{g}}_k}} \right\|^2}$. We also employ Algorithm 2 under this power allocation scheme. Besides, since Algorithm 2 can operate under any power allocation scheme, we also provide numerical results with Algorithm 2 under other power allocation schemes, such as equal power allocation, transmit power proportional to $\beta_k$ or $1/{\beta _k}$. We would like to point out here that when the transmit power is proportional to $1/{\beta _k}$, each user would achieve the same rate lower bound.

We assume that the pilot and data signals are transmitted with the same power, and ZF/ZF strategy is used during uplink and downlink data transmissions. Since RF chain constraint is not considered in \cite{Bjornson2015}, we will not include this constraint in this section.

\begin{figure}[htbp]
  \center
  \includegraphics[width=3.3in]{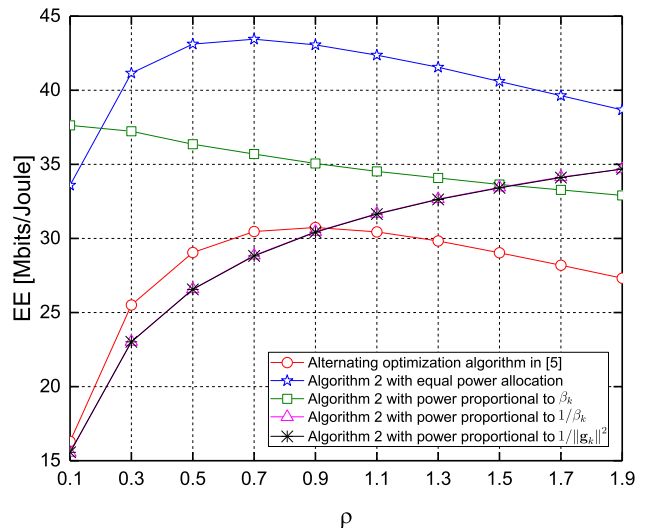}\\
  \caption{Maximum EE achieved in a single-cell massive MIMO system versus $\rho$ under perfect CSI.}\label{fig9}
\end{figure}

\begin{figure}[htbp]
  \center
  \includegraphics[width=3.3in]{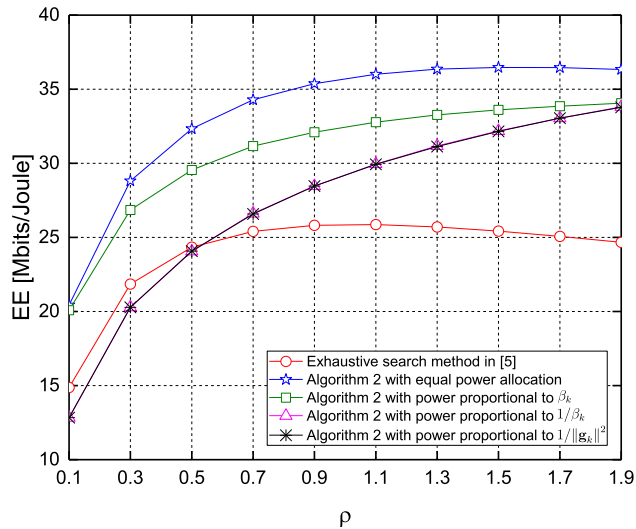}\\
  \caption{Maximum EE achieved in a single-cell massive MIMO system versus $\rho$ under imperfect CSI.}\label{fig10}
\end{figure}

Fig. \ref{fig9} and Fig. \ref{fig10} display the maximum EE achieved in a single-cell massive MIMO system versus $\rho$ (proportional to the transmit power) considering perfect and imperfect CSI, respectively. Comparing with the algorithm proposed in \cite{Bjornson2015}, Algorithm 2 can further improve the EE under both assumptions of perfect and imperfect CSI when equal power allocation scheme is used, or the transmit power is proportional to $\beta_k$. The maximum EE achieved with Algorithm 2 under transmit power proportional to $1/{\beta _k}$ overlaps with that achieved under transmit power proportional to $1/{\left\| {{{\bf{g}}_k}} \right\|^2}$, for both perfect and imperfect CSI. With equal power allocation and perfect CSI, while the curves of our algorithm and the alternating optimization algorithm in \cite{Bjornson2015} have similar shapes, Algorithm 2 improves the EE by more than 40\%, compared with the approach in \cite{Bjornson2015}. Under imperfect CSI, the authors in \cite{Bjornson2015} used an exhaustive search method. The curve with equal power allocation in Fig. \ref{fig10} shows that our algorithm can also achieve a substantial improvement, compared with the exhaustive search method in \cite{Bjornson2015}.

\section{Conclusion}
In this paper, Algorithm 2 is proposed to solve the joint uplink and downlink EE maximization problem with joint antenna selection and user scheduling in single-cell massive MIMO systems, under a limitation on the number of available RF chains. With Jensen's inequality and the power consumption model, the original joint antenna selection and user scheduling problem is converted into a combinatorial optimization problem, and we have shown that it can be solved efficiently with the developed learning-based stochastic gradient descent algorithm. We have also employed the rare event simulation method in the learning-based stochastic gradient descent algorithm to generate samples with very small probabilities. We have considered both perfect and imperfect CSI at the BS. Via numerical results,  we have provided insightful observations that are poised to be beneficial in the design of practical single-cell massive MIMO systems.

\bibliographystyle{IEEEtran}
\bibliography{antenna_selection}

\begin{thebibliography}{10}
\providecommand{\url}[1]{#1}
\csname url@samestyle\endcsname
\providecommand{\newblock}{\relax}
\providecommand{\bibinfo}[2]{#2}
\providecommand{\BIBentrySTDinterwordspacing}{\spaceskip=0pt\relax}
\providecommand{\BIBentryALTinterwordstretchfactor}{4}
\providecommand{\BIBentryALTinterwordspacing}{\spaceskip=\fontdimen2\font plus
\BIBentryALTinterwordstretchfactor\fontdimen3\font minus
  \fontdimen4\font\relax}
\providecommand{\BIBforeignlanguage}[2]{{%
\expandafter\ifx\csname l@#1\endcsname\relax
\typeout{** WARNING: IEEEtran.bst: No hyphenation pattern has been}%
\typeout{** loaded for the language `#1'. Using the pattern for}%
\typeout{** the default language instead.}%
\else
\language=\csname l@#1\endcsname
\fi
#2}}
\providecommand{\BIBdecl}{\relax}
\BIBdecl

\bibitem{Marzett2010}
T.~L. Marzetta, ``Noncooperative cellular wireless with unlimited numbers of
  base station antennas,'' \emph{IEEE Transactions on Wireless Communications},
  vol.~9, no.~11, pp. 3590--3600, November 2010.

\bibitem{Rana2010}
\BIBentryALTinterwordspacing
M.~M. Rana and M.~K. Hosain, ``Adaptive channel estimation techniques for mimo
  ofdm systems,'' \emph{International Journal of Advanced Computer Science and
  Applications}, vol.~1, no.~6, 2010. [Online]. Available:
  \url{http://dx.doi.org/10.14569/IJACSA.2010.010620}
\BIBentrySTDinterwordspacing

\bibitem{AndrewsBuzziChoiEtAl2014}
J.~G. Andrews, S.~Buzzi, W.~Choi, S.~V. Hanly, A.~Lozano, A.~C.~K. Soong, and
  J.~C. Zhang, ``What will 5{G} be?'' \emph{IEEE Journal on Selected Areas in
  Communications}, vol.~32, no.~6, pp. 1065--1082, June 2014.

\bibitem{LarssonEdforsTufvessonEtAl2014}
E.~G. Larsson, O.~Edfors, F.~Tufvesson, and T.~L. Marzetta, ``Massive {MIMO}
  for next generation wireless systems,'' \emph{IEEE Communications Magazine},
  vol.~52, no.~2, pp. 186--195, February 2014.

\bibitem{Bjornson2015}
E.~Björnson, L.~Sanguinetti, J.~Hoydis, and M.~Debbah, ``Optimal design of
  energy-efficient multi-user {MIMO} systems: Is {M}assive {MIMO} the answer?''
  \emph{IEEE Transactions on Wireless Communications}, vol.~14, no.~6, pp.
  3059--3075, June 2015.

\bibitem{LiuDuSun2017}
Z.~Liu, W.~Du, and D.~Sun, ``Energy and spectral efficiency tradeoff for
  {M}assive {MIMO} systems with transmit antenna selection,'' \emph{IEEE
  Transactions on Vehicular Technology}, vol.~66, no.~5, pp. 4453--4457, May
  2017.

\bibitem{RanaVesiloCollings2016}
M.~T.~A. Rana, R.~Vesilo, and I.~B. Collings, ``Antenna selection in {M}assive
  {MIMO} using non-central principal component analysis,'' in \emph{2016 26th
  International Telecommunication Networks and Applications Conference
  (ITNAC)}, Dec 2016, pp. 283--288.

\bibitem{BenmimouneDriouchAjibEtAl2015}
M.~Benmimoune, E.~Driouch, W.~Ajib, and D.~Massicotte, ``Feedback reduction and
  efficient antenna selection for {M}assive {MIMO} system,'' in \emph{2015 IEEE
  82nd Vehicular Technology Conference (VTC2015-Fall)}, Sept 2015, pp. 1--6.

\bibitem{QinLiLvEtAl2016}
S.~Qin, G.~Li, G.~Lv, G.~Zhang, and H.~Hui, ``${L_{{1 \mathord{\left/
  {\vphantom {1 2}} \right. \kern-\nulldelimiterspace} 2}}}$-{R}egularization
  based antenna selection for {RF}-chain limited {M}assive {MIMO} systems,'' in
  \emph{2016 IEEE 84th Vehicular Technology Conference (VTC-Fall)}, Sept 2016,
  pp. 1--5.

\bibitem{TaiChungLee2015}
T.~H. Tai, W.~H. Chung, and T.~S. Lee, ``A low complexity antenna selection
  algorithm for energy efficiency in {M}assive {MIMO} systems,'' in \emph{2015
  IEEE International Conference on Data Science and Data Intensive Systems},
  Dec 2015, pp. 284--289.

\bibitem{ArashYazdianFazelEtAl2017}
M.~Arash, E.~Yazdian, M.~S. Fazel, G.~G. de~Oliveira~Brante, and M.~A. Imran,
  ``Employing antenna selection to improve energy efficiency in {M}assive
  {MIMO} systems,'' \emph{Trans. Emerging Telecommunications Technologies},
  vol.~28, 2017.

\bibitem{GaoEdforsLiuEtAl2013}
X.~Gao, O.~Edfors, J.~Liu, and F.~Tufvesson, ``Antenna selection in measured
  {M}assive {MIMO} channels using convex optimization,'' in \emph{2013 IEEE
  Globecom Workshops (GC Wkshps)}, Dec 2013, pp. 129--134.

\bibitem{ArashYazdianFazel2016}
M.~Arash, E.~Yazdian, and M.~Fazel, ``Antenna selection: A novel approach to
  improve energy efficiency in {M}assive {MIMO} systems,'' in \emph{2016 6th
  International Conference on Computer and Knowledge Engineering (ICCKE)}, Oct
  2016, pp. 106--110.

\bibitem{Lee2013}
B.~M. Lee, J.~Choi, J.~Bang, and B.~C. Kang, ``An energy efficient antenna
  selection for large scale green {MIMO} systems,'' in \emph{2013 IEEE
  International Symposium on Circuits and Systems (ISCAS2013)}, May 2013, pp.
  950--953.

\bibitem{ElkhalilKammounAl-NaffouriEtAl2016}
K.~Elkhalil, A.~Kammoun, T.~Y. Al-Naffouri, and M.~S. Alouini, ``A blind
  antenna selection scheme for single-cell uplink {M}assive {MIMO},'' in
  \emph{2016 IEEE Globecom Workshops (GC Wkshps)}, Dec 2016, pp. 1--6.

\bibitem{LiuGaoYangEtAl2017}
H.~Liu, H.~Gao, S.~Yang, and T.~Lv, ``Low-complexity downlink user selection
  for {M}assive {MIMO} systems,'' \emph{IEEE Systems Journal}, vol.~11, no.~2,
  pp. 1072--1083, June 2017.

\bibitem{AlyamiKostanic2016}
G.~Alyami and I.~Kostanic, ``A low complexity user selection scheme with linear
  precoding for {M}assive {MIMO} systems,'' in \emph{International Journal of
  Computer Science}, 2016.

\bibitem{Maimaiti2019}
\BIBentryALTinterwordspacing
S.~Maimaiti, G.~Chuai, W.~Gao, K.~Zhang, X.~Liu, and Z.~Si, ``A low-complexity
  algorithm for the joint antenna selection and user scheduling in multi-cell
  multi-user downlink massive mimo systems,'' \emph{EURASIP Journal on Wireless
  Communications and Networking}, vol. 2019, no.~1, p. 208, 2019. [Online].
  Available: \url{https://doi.org/10.1186/s13638-019-1529-7}
\BIBentrySTDinterwordspacing

\bibitem{DongTangShenzhen2017}
Y.~Dong, Y.~Tang, and K.~Z. Shenzhen, ``Improved joint antenna selection and
  user scheduling for {M}assive {MIMO} systems,'' in \emph{2017 IEEE/ACIS 16th
  International Conference on Computer and Information Science (ICIS)}, May
  2017, pp. 69--74.

\bibitem{BenmimouneDriouchAjibEtAl2015a}
M.~Benmimoune, E.~Driouch, W.~Ajib, and D.~Massicotte, ``Joint transmit antenna
  selection and user scheduling for {M}assive {MIMO} systems,'' in \emph{2015
  IEEE Wireless Communications and Networking Conference (WCNC)}, March 2015,
  pp. 381--386.

\bibitem{XuLiuJiangEtAl2014}
G.~Xu, A.~Liu, W.~Jiang, H.~Xiang, and W.~Luo, ``Joint user scheduling and
  antenna selection in distributed {M}assive {MIMO} systems with limited
  backhaul capacity,'' \emph{China Communications}, vol.~11, no.~5, pp. 17--30,
  May 2014.

\bibitem{Rana2017}
M.~{Rana}, L.~{Li}, and S.~W. {Su}, ``Distributed state estimation over
  unreliable communication networks with an application to smart grids,''
  \emph{IEEE Transactions on Green Communications and Networking}, vol.~1,
  no.~1, pp. 89--96, March 2017.

\bibitem{PillaiSuelSeunghunCha2005}
S.~U. {Pillai}, T.~{Suel}, and {Seunghun Cha}, ``The perron-frobenius theorem:
  some of its applications,'' \emph{IEEE Signal Processing Magazine}, vol.~22,
  no.~2, pp. 62--75, March 2005.

\bibitem{Berny2001}
A.~Berny, \emph{Statistical Machine Learning and Combinatorial
  Optimization}.\hskip 1em plus 0.5em minus 0.4em\relax Berlin, Heidelberg:
  Springer Berlin Heidelberg, 2001, pp. 287--306.

\bibitem{JiangZhangRenEtAl2017}
C.~Jiang, H.~Zhang, Y.~Ren, Z.~Han, K.~C. Chen, and L.~Hanzo, ``Machine
  learning paradigms for next-generation wireless networks,'' \emph{IEEE
  Wireless Communications}, vol.~24, no.~2, pp. 98--105, April 2017.

\bibitem{Shi2018}
J.~{Shi}, W.~{Wang}, J.~{Wang}, and X.~{Gao}, ``Machine learning assisted
  user-scheduling method for massive mimo system,'' in \emph{2018 10th
  International Conference on Wireless Communications and Signal Processing
  (WCSP)}, Oct 2018, pp. 1--6.

\bibitem{AuBeck2001}
S.-K. Au and J.~L. Beck, ``Estimation of small failure probabilities in high
  dimensions by subset simulation,'' \emph{Probabilistic Engineering
  Mechanics}, vol.~16, no.~4, pp. 263 -- 277, 2001.

\bibitem{RusekPerssonLauEtAl2013}
F.~Rusek, D.~Persson, B.~K. Lau, E.~G. Larsson, T.~L. Marzetta, O.~Edfors, and
  F.~Tufvesson, ``Scaling up {MIMO}: Opportunities and challenges with very
  large arrays,'' \emph{IEEE Signal Processing Magazine}, vol.~30, no.~1, pp.
  40--60, Jan 2013.

\bibitem{Ngo2013}
H.~Q. Ngo, E.~G. Larsson, and T.~L. Marzetta, ``Energy and spectral efficiency
  of very large multiuser {MIMO} systems,'' \emph{IEEE Transactions on
  Communications}, vol.~61, no.~4, pp. 1436--1449, April 2013.

\bibitem{BjornsonOttersten2010}
E.~{Bjornson} and B.~{Ottersten}, ``A framework for training-based estimation
  in arbitrarily correlated rician mimo channels with rician disturbance,''
  \emph{IEEE Transactions on Signal Processing}, vol.~58, no.~3, pp.
  1807--1820, March 2010.

\bibitem{Zuev2021}
K.~M. Zuev, \emph{Subset Simulation Method for Rare Event Estimation: An
  Introduction}.\hskip 1em plus 0.5em minus 0.4em\relax Berlin, Heidelberg:
  Springer Berlin Heidelberg, 2021, pp. 1--25.

\bibitem{Berny2000}
A.~Berny, ``Selection and reinforcement learning for combinatorial
  optimization,'' in \emph{International Conference on Parallel Problem Solving
  from Nature}.\hskip 1em plus 0.5em minus 0.4em\relax Springer, 2000, pp.
  601--610.

\bibitem{Gallagher2005}
M.~{Gallagher} and M.~{Frean}, ``Population-based continuous optimization,
  probabilistic modelling and mean shift,'' \emph{Evolutionary Computation},
  vol.~13, no.~1, pp. 29--42, March 2005.

\bibitem{Zhang2019}
K.~Zhang, A.~Koppel, H.~Zhu, and T.~Basar, ``Global convergence of policy
  gradient methods to (almost) locally optimal policies,'' 2019.

\end{thebibliography}

\end{document}